\documentclass[aps,pre,11pt,onecolumn,notitlepage,superscriptaddress,floatfix]{revtex4-2}

\usepackage{epsfig}
\usepackage{amssymb,amsmath}
\usepackage[mathlines]{lineno}
\usepackage{etoolbox} 
\usepackage{array}
\usepackage{graphicx,color,subfigure,bm,amsmath,amssymb}
\usepackage{hyperref}
\usepackage[mathlines]{lineno}
\hypersetup{
    colorlinks=true,
    linkcolor=cyan,
    filecolor=magenta,      
    urlcolor=blue,
    citecolor=blue,
}
\usepackage{graphicx}
\usepackage{dcolumn}
\usepackage{bm}
\usepackage{color}
\usepackage{siunitx}
\let\oldequation\equation
\let\oldendequation\endequation

\renewenvironment{equation}
  {\linenomathNonumbers\oldequation}
  {\oldendequation\endlinenomath}

\begin{document}

\title{Biophysical Fluid Dynamics in a Petri Dish}

\author{George T. Fortune}
\email[]{gtf22@damtp.cam.ac.uk}
\author{Eric Lauga}
\email[]{e.lauga@damtp.cam.ac.uk}
\author{Raymond E Goldstein}
\email[]{R.E.Goldstein@damtp.cam.ac.uk}
\affiliation{Department of Applied Mathematics and Theoretical Physics, Centre for Mathematical Sciences, 
University of Cambridge, Wilberforce Road, Cambridge, CB3 0WA, United Kingdom}

\date{\today}
             
\begin{abstract}
The humble Petri dish is perhaps the simplest setting in which to examine the locomotion of swimming 
organisms, particularly those whose body size is tens of microns to millimetres. The fluid layer in 
such a container has a bottom no-slip surface and a stress-free upper boundary. It is of fundamental 
interest to understand the flow fields produced by 
the elementary and composite singularities of Stokes flow in
this geometry.
Building on the few particular cases that have 
previously been considered in the literature, 
we study here the image systems for the primary singularities of Stokes flow subject to such 
boundary conditions \textemdash the stokeslet, rotlet, source, rotlet dipole, source dipole and stresslet
\textemdash paying particular attention to the far-field behavior. In several key situations, 
the depth-averaged fluid flow is accurately captured by the solution of an associated Brinkman equation 
whose screening length is proportional to the depth of the fluid layer.  The case of hydrodynamic bound states 
formed by spinning microswimmers near 
a no-slip surface, discovered first using the alga {\it Volvox},
is reconsidered in
the geometry of a Petri dish, where the power-law  
attractive interaction between
microswimmers acquires unusual exponentially screened oscillations.
\end{abstract}
\maketitle

\section{Introduction}
Since its development in 1887 by the German physician Julius Petri \cite{Petri87} for the facilitation of 
cell culturing, extending the bacterial culture methods pioneered by his mentor Robert Koch \cite{Koch81}, 
the Petri dish has become an integral part of any biology laboratory. While still primarily 
used for culturing cells, providing storage space whilst reducing the risk of contamination, its simplicity 
and functionality allows it to be used in a wide range of other contexts: in chemistry to dry out 
precipitates and evaporate solvents (e.g.~when studying Liesegang rings
\cite{Henisch88,Lagzi04}) or in entomology where they are convenient enclosures to study the behaviour 
of insects and small animals \cite{Franks22,Valente07}. A Petri dish 
environment is also a simple and common setting in which to examine the locomotion of swimming 
organisms, particularly those whose body size is tens of microns to 
millimetres \cite{Bentley21,Dunkel13b,Drescher11,Bazazi12,Zaki21}. The boundary condition at the bottom surface of such a container can be approximated as 
no-slip, while the top of the fluid is 
stress-free. Hence, a general question is: how does confinement in a Petri dish alter the nature of the flow induced by motile organisms? 

The framework to answer this question lies of course with Green's functions.
In low Reynolds number fluid mechanics governed by the Stokes equations \cite{Stokes51}, the 
most important such function corresponds to the flow induced by a point force in an unbounded fluid 
and decays as $1/r$. First written down by Lorentz \cite{Lorentz96} and later denoted a 
Stokeslet \cite{Hancock53}, it has been used to solve a wide range of fluid dynamical problems 
(see Happel and Brenner \cite{Happel12} and Kim and Karrila \cite{Kim05} for general overviews). One powerful extension to the Stokeslet 
involves a multipole expansion similar to that in electrostatics. The fluid flow caused by the 
motion of an arbitrary rigid body through a viscous fluid can be represented as that from a collection of point 
forces at the surface of the body \cite{Kim05}. Expanding the Stokeslet produced at 
an arbitrary point on the body's surface as a Taylor series about the center of the body and then summing 
these contributions in the far field, one obtains a perturbation expansion for the 
fluid flow induced by the body \cite{Chwang75}. Regardless of the particular shape of the particle, 
the fluid velocity field will exhibit common features. The leading order $1/r$ term is still a Stokeslet,
but at higher orders, one finds distinct singularities. In particular the $1/r^2$ term, denoted a 
force dipole, can be separated into a symmetric part, denoted a stresslet \cite{Batchelor70}, that 
corresponds to a symmetric hydrodynamic stress applied locally to the fluid, and an anti-symmetric part, 
denoted a rotlet \cite{Chwang71} (called a couplet by Batchelor \cite{Batchelor70})), corresponding to a local hydrodynamic
torque that produces rotational motion.

\begin{figure*}
\centering\includegraphics[trim={0 0cm 0 0cm}, clip, width=1.00\textwidth]{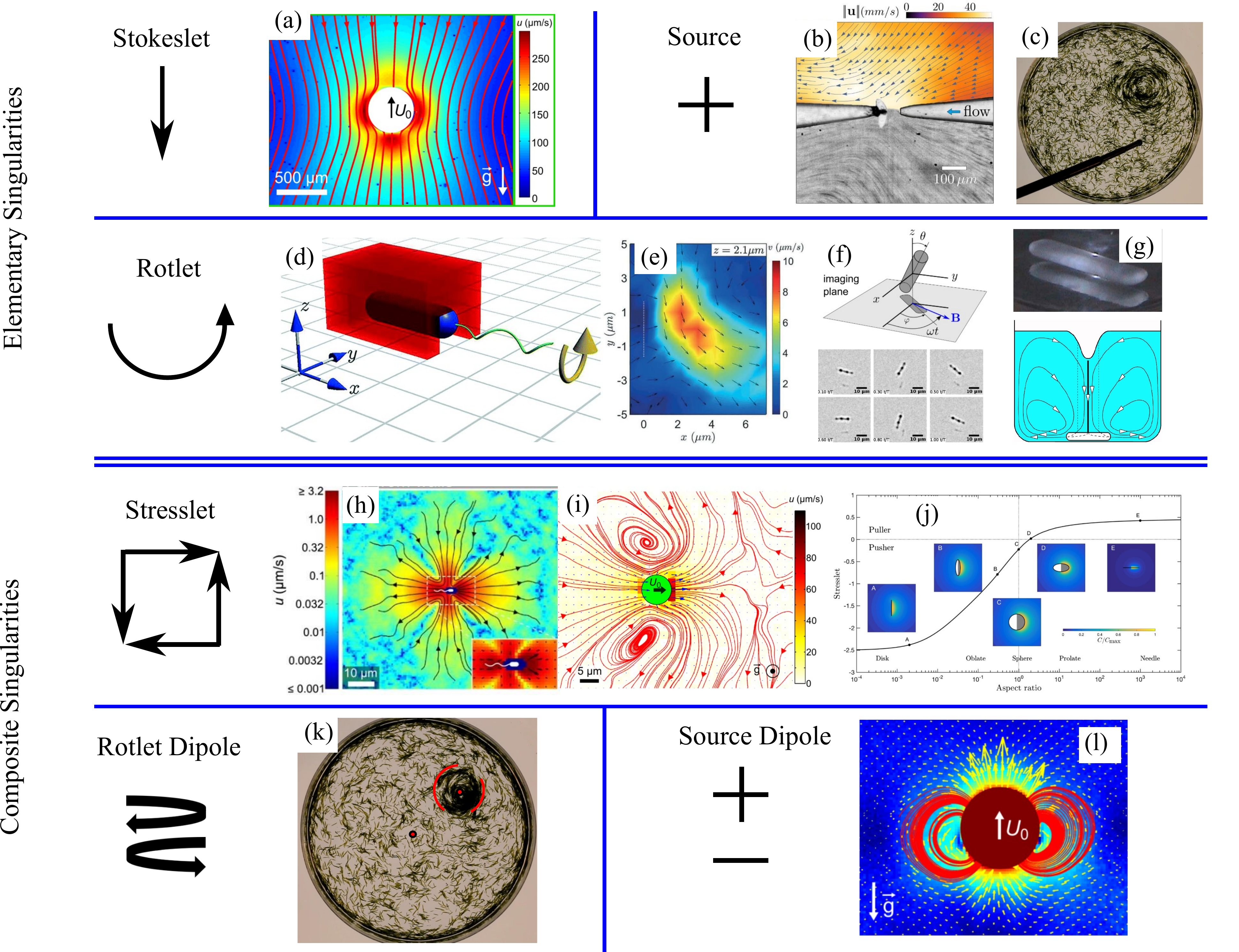}
\caption{Stokes singularities in biological fluid mechanics. [a-g] Elementary singularities. Stokeslet flow is found in (a) far-field flow around \textit{Volvox carteri} \cite{Drescher10}.  Source flows arise from injection of fluid from 
a micropipette into a Petri dish 
in studies of (b) dinoflagellates \cite{Jalaal20} and (c) plant-animal worms \cite{Fortune21}. 
Rotlet flows arise from (d) the bacterium \textit{Escherichia coli} under confinement, 
generating flow
field in (e) \cite{Gao15}, (f) a magnetic nano stir bar \cite{Gires20}, and (g) a 
macroscopic stirrer \cite{Halasz07}. [h-l] Composite singularities. Stresslets 
arise from (h) the pusher \textit{E. coli} \cite{Drescher11}, 
(i) the puller alga \textit{Chlamydomonas reinhardtii} \cite{Drescher10}, and (j) a 
phoretic Janus particle that changes from pusher to puller 
as a function of its aspect ratio \cite{Lauga16b}.  A rotlet dipole flow is induced
by (k) a circular mill of \textit{Symsagittifera roscoffensis} \cite{Fortune22}. 
A source is found in (l) the near-field flow induced by \textit{Volvox carteri} 
after the Stokeslet contribution is subtracted \cite{Drescher10}.
}
\label{fig1}
\end{figure*}

A well chosen distribution of such Stokes singularities that exploits the inherent symmetries of the 
system in question can be used to solve Stokes equations in a wide range of geometries and biological 
contexts \cite{Kim05}. Figure \ref{fig1} illustrates the breadth of this approach, giving 
examples of biological flows associated with each of the low order Stokes singularities. Although 
classically in biological fluid dynamics the stresslet is the most common Stokes singularity considered \cite{Lauga20}, 
one sees that all low order Stokes singularities arise in familiar contexts.  

The key question addressed here is thus: what is the fluid flow resulting from any Stokes singularity 
placed in a fluid layer between a rigid lower no-slip  boundary and an upper stress-free surface. 
Although a few cases have been investigated in the literature, there has not been a systematic breakdown of the possible cases that arise. This was first considered by Liron and Mochon \cite{Liron76} who derived an exact solution in integral form for a Stokeslet. Subsequent  work on this problem includes a theoretical study of bacterial swarms on agar \cite{Dauparas16}, which contained a calculation of the leading order far field contribution to the flow from both a Stokeslet and a Rotlet when placed in a Petri dish configuration. This was further developed by Mathijssen, et. al. \cite{Mathijssen16}, who derived a numerically tractable approximation for the flow field produced by a Stokeslet and hence the flow field produced by a force- and torque-free micro-swimmer in a Petri dish. 

In this paper, paying particular attention to the 
far-field behavior, we systematically extend and generalize these works beyond Stokeslets by computing exact 
expressions for the flow components $u_j$ generated in a Petri dish of height $H$ by the biologically 
relevant low-order primary and composite
singularities of Stokes flow:
\begin{flalign}
   &\text{1. The Stokeslet: } u^{k}_j = \lambda_{F} 
\left(\delta_{jk}/r + x_jx_k/r^3 \right), \label{eq:stokesletdef} \\
&\text{2. Rotlet: } u^k_j = \lambda_{R} \epsilon_{jkp}x_p/r^3, &&\label{eq:rotletdef} \\
&\text {3. Source: } u_j = \lambda_S x_j / r^3, &&\label{eq:sourcedef} \\
&\text{4. General stresslet: } u_j^{k, \, l} = \lambda_{C} x_jx_kx_l/r^5,&& \label{eq:stressletdef} \\
&\text{5. Rotlet dipole: } u_j^k = \lambda_{RD} \epsilon_{jpk}x_k x_p/r^5, &&\label{eq:rotletdipoledef} \\
&\text{6. Source dipole: } 
u_j^k = \lambda_{SD}\left(\delta_{jk}/r^3 - 3x_jx_k/r^5 \right).&& \label{eq:sourcedipoledef}
\end{flalign}

\begin{table}[t]
	\begin{center}
		\small
		\begin{tabular}{|c|c|c|c|c|}
  \hline
			Singularity &  & Location & Exact  & Far field \\
   			 &  &  & solution & approx. \\
   \hline
   \hline
   Source & $+$ & Main text & eq. 23 & eqs. 29-30 \\
   \hline
			Stokeslet & $\downarrow$& Main text & eq. 24 & eqs. 32-34.\\
   \hline
   Rotlet & $\circlearrowleft$&  Appendix B & eq. B6 & eqs. B8-B12\\
   \hline
Stresslet & $\downarrow _{\Big{.\leftarrow}}^{\Big{.\rightarrow}}\uparrow$ & Appendix C & eq. C4 & eqs. C7-C9\\
   \hline
   Rotlet dipole  &$_{\Big{.\circlearrowleft}}^{\Big{.\circlearrowright}}$ & Appendix D & eq. D9 &  eqs. D13-D17\\
   \hline
   Source dipole & $\pm$ & Appendix E & eq. E4 & eqs. E6-8\\
   \hline

		\end{tabular}
		\normalsize
		\caption{Location of results for various singularities.}
		\label{table:1}
	\end{center}
\end{table}

\begin{figure}[t]
\centering\includegraphics[trim={0 0cm 0 0cm}, clip, width=0.7\columnwidth]{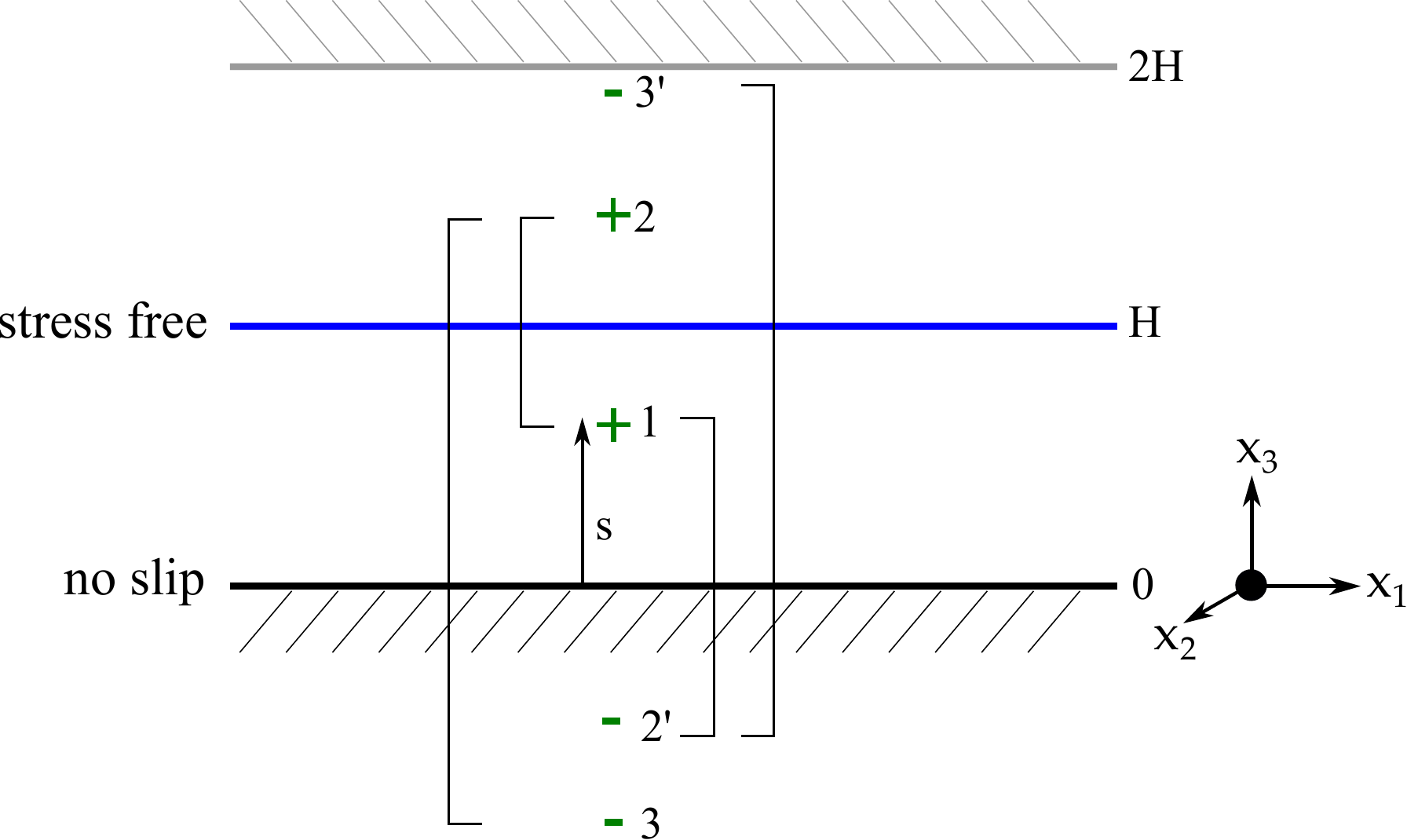}
\caption{Stokes singularity in a Petri dish. The positive singularity
is located at $z=s$ and labelled $1$. Its reflection across the
no-stress surface at $z=H$ is labelled $2$ and across the no-slip surface
at $z=0$ is $2'$, and so on. An alternate approach uses the full 
solution for a single no-slip surface and extends the domain to include a no-slip surface 
at $z=2H$.}
\label{fig2}
\end{figure}

Note that here, $j,k$ and $l$ are free indices while the $\lambda_{i}$ are dimensional constants denoting 
the strength of the singularities, with dimensions \si{\metre^2\second^{-1}} for the 
Stokeslet, \si{\metre^3\second^{-1}} for the rotlet, source and stresslet and \si{\metre^4\second^{-1}} 
for the rotlet dipole and source dipole. For clarity, we only present in the main text analysis for a 
source and a Stokeslet, namely the simplest and the most common singularity respectively. The results 
for the other singularities are given in Appendices \ref{rotletappendix}-\ref{sourcedipoleappendix}. 
Table \ref{table:1} lists the locations of all these results in the paper.
We adopt the geometry 
of Fig.~\ref{fig2}, with in-plane
coordinates 
$(x_1,x_2)$, the no-slip surface at $x_3=0$ and the stress-free
surface at $x_3=H$. 

In \S\ref{repeatedreflection}, we calculate for both a source and a Stokeslet a particular solution to the Stokes equations 
generated by summing the infinite image system of Stokes singularities that is formed by repeatedly reflecting 
the initial singularity in both of the vertical boundaries. Then in \S\ref{auxiliary_solution}, an 
auxiliary solution is calculated using a Fourier transform method so that the sum of the two solutions 
is an exact solution for the full boundary conditions. In \S\ref{farfieldsolution}, a contour 
integral approach is used to calculate the leading order term of the fluid velocity
in the far-field of a source. 

This methodology, applied to both the source and the Stokeslet 
in \S\ref{auxiliary_solution}-\ref{farfieldsolution}, is applied to the rest of the most commonly used Stokes 
singularities, (namely a rotlet, a general stresslet, a rotlet dipole and a source dipole), in 
Appendices \ref{rotletappendix}-\ref{sourcedipoleappendix}. 
Finally, as an application of these results, \S\ref{boundstates} reconsiders in the geometry of the 
Petri dish the problem of hydrodynamic bound states, 
first discovered using the
green alga {\it Volvox} near a no-slip surface 
\cite{Drescher09} and later rediscovered in multiple contexts.
The concluding \S\ref{discussion} summarises the main
results of the paper.  

In particular, we note that higher order in-plane Stokes singularities can be found by 
differentiating the solutions with respect to a horizontal coordinate $x_\alpha$. 
Since all other Stokes singularities can be expressed in terms of derivatives of these singularities, we conclude that the leading order contribution to the fluid velocity in the far field 
for an arbitrary Stokes singularity is separable in $x_3$, 
either decaying exponentially radially or having $x_3$ dependence of the form $x_3 (1 - x_3/2H)$. Hence, 
for many situations where the forcing can be 
modelled as a sum of Stokes singularities, the depth-averaged fluid 
flow can be captured by an associated Brinkman equation with a screening length proportional to $H$.

\section{Singularity in a Petri dish} \label{singularity}

Consider, as in Fig.~\ref{fig2}, a Stokes singularity $f$, located at the point 
$(x_1, x_2, x_3) = (0, 0, s)$ between a rigid 
lower surface at $x_3 = 0$ and an upper free surface at $x_3 = H$, which generates a fluid flow 
$\boldsymbol{u} = (u_1,u_2,u_3)$. At $x_3 = 0$, we impose the no-slip boundary conditions 
\begin{equation}
u_1 = u_2 = u_3 = 0. 
\end{equation}
The capillary length $\lambda_{\text{cap}}$ for a water-air interface is
$\lambda_{\text{cap}} = \sqrt{\gamma_{w}/\rho_{w}g} \approx \SI{2.73}{\milli\metre}$,
where $\rho_w=\SI{997}{\kilogram\metre^{-3}}$ is the density of water, 
$\gamma_{w}$ as $\SI{72.8}{\milli\newton \metre^{-1}}$ 
is the air-water surface tension, and $g=\SI{9.81}{\metre \second^{-2}}$ is the gravitational acceleration. 
Since in a Petri dish 
$\lambda_{\text{cap}}$ and $H$ are similar in size, at the free surface, surface tension and gravitational effects 
are of similar magnitudes. Together, they restrict the vertical deformation of the interface. Hence, 
we assume the limit of no deformation in the vertical direction, fixing $H$ as a constant. The 
self-consistency of this assumption is explored later in \S\ref{discussion}. The dynamic boundary 
condition $u_3 = DH/Dt$ thus simplifies to 
\begin{equation}
u_3 = 0 \quad \text{at} \quad x_3 = H. 
\end{equation}
A force balance at $x_3=H$, $\sigma_{x_1x_3} = 
\sigma_{x_2x_3} = 0$, implies
\begin{equation}
\frac{\partial u_1}{\partial x_3} = \frac{\partial u_2}{\partial x_3} = 0 \quad \text{at}  \quad x_3 = H. 
\end{equation}

We nondimensionalize this system, scaling lengths with $H$ and velocities with $U_S$, where for a 
singularity of strength $\lambda_{S}$ that decays in the far field like $1/r^{n}$, 
$U_S = \lambda_S H^{-n}$. For notational simplicity, we define 
\begin{subequations}
\begin{align}
(x, \, y, \, z) &= (x_1, \, x_2, \, x_3)/H, \\
(u_x, \, u_y, \, u_z) &= (u_1, \, u_2, \, u_3)/U_S, \\ 
h &= s/H. 
\end{align}
\end{subequations}
The boundary conditions become
\begin{subequations}
\begin{align}
    u_{x} &= u_{y} = u_{z} = 0 \text{ at } z = 0, \\
    \frac{\partial u_{x}}{\partial z} &= \frac{\partial u_{y}}{\partial z} = u_{z} = 0 \text{ at } z = 1.
\end{align}
\label{final_bcs}
\end{subequations}

\section{Repeated Reflection Solution} \label{repeatedreflection}
We first examine the extent to which we can satisfy these boundary conditions through a 
distribution of image singularities. Following the canonical approach of Liron and Mochon \cite{Liron76}, 
for a singularity placed at $x_3 = s$ [the green ${\textbf{+}}$ 
labelled $1$ in Fig.~\ref{fig2}], placing an image singularity of the same sign at $x_3 = 2H-s$ 
(label 2) satisfies the free surface boundary condition at $x_3=H$. Similarly, placing 
an image singularity of the opposite sign at $x_3 = -s$ ($2'$) 
partially satisfies the 
no-slip boundary condition, but singularity $2$ fails the no-slip boundary condition and thus must be reflected 
about $x_3 = 0$, changing its sign at location $3$, Similarly, singularity $2'$ fails the 
free surface boundary condition and thus must be reflected in $x_3=H$ to give singularity $3'$. Repeating this \textit{ad infinitum}, namely inverting the sign when reflecting in 
the no-slip $x_3 = 0$ boundary and keeping the same sign when reflecting in the free surface $x_3 = H$ boundary, 
gives an infinite series of singularities that constitutes the repeated reflection solution for that 
singularity. 

In rescaled units, if we define the singularity locations
$\boldsymbol{r}_{1n} = (x \, , \, y \, , \, z - h + 4n)$, 
$\boldsymbol{r}_{2n} = (x \, , \, y \, , \, z - h + (4n+2))$,
$\boldsymbol{R}_{1n} = (x \, , \, y \, , \, z + h + 4n)$, and
$\boldsymbol{R}_{2n} = (x \, , \, y \, , \, z + h + (4n+2))$,
then the repeated reflection solution is but one case of the general function 
$\mathcal{L}(f)$ for an arbitrary function $f$, 
\begin{equation}
        \mathcal{L} = \sum_{n = -\infty}^{\infty}\!\!\!\! \left\{f(\boldsymbol{r}_{1n})
        - f(\boldsymbol{r}_{2n}) 
        - f(\boldsymbol{R}_{1n}) + f(\boldsymbol{R}_{2n}) \right\}.
        \label{eq:intialsummation}
\end{equation}
While intuitive, this series expansion is unwieldy. For the particular case $f = 1/r$, 
a Bessel function identity can be used to obtain the
integral form
\begin{equation}
     \mathcal{L}\left(\frac{1}{r}\right) = \int_0^\infty\!\! d\lambda\, \frac{2J_0(\lambda\rho)}{\cosh(\lambda )}\left\{
    \begin{array}{ll}
     \sinh h\lambda \, \cosh(1-z)\lambda, \\
     \sinh z\lambda \, \cosh(1-h)\lambda, 
     \end{array} \right.  \label{eq:repeatedreflection1r} 
\end{equation}
where $\rho = \sqrt{x^2 + y^2}$ and here and below the upper expression holds for $z>h$ and the lower 
for $z<h$. 
Higher order solutions are obtained from this result through algebraic manipulation,
as shown in Appendix \ref{appendixnotation} for the third and fifth order cases. 
From those results, we find the repeated reflection solution $v_j$ for a source $x_j/r^3$,
\begin{align}
    v_j =& \delta_{j \alpha}x_{\alpha} \mathcal{L}\left( \frac{1}{r^3} \right) 
    + \delta_{j 3} \mathcal{L} \left( \frac{z}{r^3} \right) \nonumber \\
    =& \frac{2 x_{\alpha}}{\rho} \delta_{j \alpha} \int^{\infty}_{0} \lambda d\lambda \frac{J_1(\lambda \rho)}{\cosh{\lambda }}\left\{
    \begin{array}{ll}
     \sinh h\lambda \, \cosh(1-z)\lambda,  \\
     \sinh z\lambda \, \cosh(1-h)\lambda, 
     \end{array} \right. \nonumber\\ 
     &+ 2 \delta_{j 3} \int^{\infty}_0 \lambda d\lambda \frac{J_0(\lambda \rho)}{\cosh{\lambda}} \left\{
    \begin{array}{ll}
     \sinh h\lambda \, \sinh(1-z)\lambda,  \\
     -\cosh z\lambda \, \cosh(1-h)\lambda.
     \end{array} \right.
     \label{auxiliarysolution}
\end{align}
Similarly, for a Stokeslet $\delta_{jk}/r + x_jx_k/r^3$, we find 
\begin{align}
    v_j^k =& \delta_{jk} \mathcal{L}\left( \frac{1}{r} \right) +  \delta_{j\alpha}\delta_{k\beta} x_{\alpha}x_{\beta}\mathcal{L}{\left( \frac{1}{r^3} \right)} 
    + (\delta_{j\alpha}\delta_{k3} + \delta_{k\alpha}\delta_{j3})x_{\alpha}\mathcal{L}\left( \frac{z}{r^3} \right) + \delta_{j3}\delta_{k3}\mathcal{L}\left( \frac{z^2}{r^3} \right) \nonumber\\
    =& 2\left( \delta_{jk} + \delta_{j3}\delta_{k3} \right) \int^{\infty}_{0} d\lambda \, \frac{J_0(\lambda \rho)}{\cosh{\lambda}} \left\{
    \begin{array}{ll}
     \sinh h\lambda \, \cosh(1-z)\lambda, \\
     \sinh z\lambda \, \cosh(1-h)\lambda,
     \end{array} \right. \nonumber \\
    &+ 2 \Bigg{(} \frac{x_{\alpha}x_{\beta}}{\rho}\delta_{j\alpha}\delta_{k\beta} - \rho\delta_{j3}\delta_{k3} \Bigg{)} \int^{\infty}_{0} \lambda d\lambda \, \frac{J_1(\lambda \rho)}{\cosh{\lambda}} 
    \times \left\{
    \begin{array}{ll}
     \sinh h\lambda \, \cosh(1-z)\lambda, \\
     \sinh z\lambda \, \cosh(1-h)\lambda,
     \end{array} \right. \nonumber \\
    &+ 2x_{\alpha}\Big{(} \delta_{j3}\delta_{k\alpha} + \delta_{k3}\delta_{j\alpha} \Big{)} \int^{\infty}_{0} \lambda d\lambda \, \frac{J_0(\lambda \rho)}{\cosh{\lambda}}
    \times \left\{
    \begin{array}{ll}
     \sinh h\lambda \, \sinh(1-z)\lambda, \\
     -\cosh z\lambda \, \cosh(1-h)\lambda,
     \end{array} \right.
\end{align}
Similar expressions can be constructed for the other commonly used Stokes singularities 
(see Appendix \ref{rotletappendix} for the rotlet, \ref{stressletappendix} for the stresslet, \ref{rotletdipoleappendix} for the rotlet dipole, and \ref{sourcedipoleappendix} for the source dipole).

These results obtained via the repeated reflection solution can also be found directly from 
Liron's solution \cite{Liron76} for a point
force between two no-slip walls by setting the separation
in that calculation to be $2H$, placing a second 
force at $2H-s$ and observing that the  
reflection symmetry of the problem about the midline at 
$x_3=H$ guarantees a stress-free condition at the
midline.

Due to the nature of the algebraic manipulations performed above, these integral expressions 
do not converge when in the horizontal plane of the singularity $x_3 = s$. Instead, it transpires that the 
correct integral expression to use instead is $\left(v_j^k\big{|}_{x_3 \rightarrow s^+} 
+ v_j^k \big{|}_{x_3 \rightarrow s^-} \right)/2$, the average of the integrals as $x_3$ tends to 
$s$ from both directions. 

\section{Auxiliary Solution} 
\label{auxiliary_solution}

In a scalar problem, such as a set of electric charges, the repeated reflection solution would 
solve the full system. However, our singularities are vectors and thus the repeated reflection solution does not 
satisfy all the boundary conditions. If we write the full fluid velocity field $u^k_j$ as 
$u^k_j = v^k_j + w^k_j$, then the auxiliary solution $w^k_j$ satisfies
\begin{equation}
    \mu \nabla^2 w_j = \frac{\partial q}{\partial x_j} \quad , \quad \frac{\partial w_j}{\partial x_j} = 0 \quad \longrightarrow \quad \nabla^2 q = 0, \label{eq:wequations}
\end{equation}
for suitable effective pressure $q$, with boundary conditions
\begin{equation}
    w_j \Big{|}_{z = 0} = - v_j \Big{|}_{z = 0}, \quad w_3 \Big{|}_{z = 1} = - v_3 \Big{|}_{z = 1}, \quad
    \frac{\partial w_{\alpha}}{\partial z} \Bigg{|}_{z = 1} = -\frac{\partial v_{\alpha}}{\partial z} \Bigg{|}_{z = 1}, \label{eq:wbc}
\end{equation}
where $\alpha \in [1,2]$ and $j \in [1,3]$. 
For a source these are
\begin{equation}
    w_{\alpha} \Big{|}_{z = 0} = w_{3} \Big{|}_{z = 1} = \frac{\partial w_{\alpha}}{\partial z} \Bigg{|}_{z = 1} = 0, \quad
    w_3 \Big{|}_{z = 0} = 2\int^{\infty}_{0} \lambda d\lambda \frac{J_0(\lambda \rho)}{\cosh{\lambda }} \cosh{(1 - h)\lambda}. \label{eq:wsource2}
\end{equation}
Similarly for a Stokeslet, applying standard Bessel function identities, the auxiliary boundary conditions become
\begin{align}
    w_j^k \Big{|}_{z = 0} &= 2x_{\alpha} \left( \delta_{j 3} \delta_{k \alpha} + \delta_{k 3} \delta_{j \alpha} \right) 
    \int^{\infty}_{0} \lambda d\lambda \, J_0(\lambda \rho)\frac{\cosh{(1 - h)\lambda}}{\cosh{\lambda }}, \\
    \frac{\partial w_{\alpha}^k}{\partial z}\Bigg{|}_{z = H} &= 2x_{\alpha} \delta_{k 3} \int^{\infty}_{0} \lambda d\lambda \, J_0(\lambda \rho) \frac{\lambda \sinh{h\lambda}}{\cosh{\lambda }}, \\
    w_3^k \Big{|}_{z = H} &= 2\delta_{k3}\int^{\infty}_{0} \lambda d\lambda \, J_0(\lambda\rho) 
    \left( \frac{\partial}{\partial \lambda}\left( \frac{\sinh{h\lambda}}{\cosh{\lambda }} \right) - \frac{\sinh{h\lambda}}{\lambda \cosh{\lambda }} \right). \label{eq:Stokesletbcwk3}
\end{align}

We solve for $w_j$ by taking the two dimensional Fourier transform of this system with 
respect to $(x, y)$, (namely $w_j(x, y, z) \Longrightarrow \hat{w}_j(k_1, k_2, z)$), to arrive at
\begin{subequations}
\begin{align}
    \mu \left(\frac{\partial^2 \hat{w}_j}{\partial z^2} - k^2 \hat{w}_j \right) &= \delta_{j 3} \frac{\partial \hat{q}}{\partial z} + i \delta_{\alpha j}k_{\alpha}\hat{q}, \label{eq:stokes1} \\
    \frac{\partial \hat{w}_3}{\partial z} + i k_{\alpha}\hat{w}_{\alpha} & = 0, \label{eq:continuity1} \\
    \frac{\partial^2 \hat{q}}{\partial z^2} - k^2 \hat{q} &= 0, \label{eq:pressure1}
\end{align}
\end{subequations}
where $\alpha \in [1, 2]$ and $k^2 = k^2_1 + k^2_2$. From inspection, this has the general solution 
\begin{subequations}
\begin{align}
    \hat{q} &= B(k) \sinh{k(1 - z)} + C(k) \cosh{k(1 - z)}, \label{eq:qtransformedform} \\
    2\mu \hat{w}_j &= B_j(k) \sinh{k(1 - z)} + C_j(k) \cosh{k(1 - z)} \nonumber \\
    &+ (z - 1) \cosh{k(1 - z)}\left( \delta_{j 3} C - \delta_{\alpha j}\frac{i k_{\alpha}}{k}B \right) + z \sinh{k (1 - z)} \left( \delta_{j 3} B - \delta_{\alpha j}\frac{i k_{\alpha}}{k}C \right),
    \label{eq:stokeslettransformedform}
\end{align}
\end{subequations}
where $\{ B, \, C, \, B_j, \, C_j \}$, with $j \in [1,2,3]$, are independent of $z$.  From the continuity equation (\ref{eq:continuity1}) they
satisfy
\begin{align}
    C &= k B_3 + k  B - i k_{1} C_{1} - i k_{2} C_{2}, \nonumber \\
  B &= k C_3 - k  C - i k_{1} B_{1} - i k_{2} B_{2}~.
\end{align}

These constants are found on a case by case basis by transforming the boundary conditions given in (\ref{eq:wbc}) and solving through matrix methods the resulting set of eight coupled simultaneous equations in terms of $\{ k, \, h \}$. For a source, \eqref{eq:wsource2} transforms to give 
\begin{equation}
    \hat{w}_{\alpha} \Big{|}_{z = 0} = \hat{w}_3 \Big{|}_{z = 1} = \frac{\partial \hat{w}_{\alpha}}{\partial z} \Bigg{|}_{z = 1} = 0, \quad
    \hat{w}_3 \Big{|}_{z = 0} = 4 \pi \frac{\cosh{(1 - h)k}}{\cosh{k }},
\end{equation}
with corresponding full solution for $\hat{w}_j$
\begin{subequations}
\begin{align}
    \hat{w}_3 =& \frac{4\pi \cosh{(1 - h)k}}{\cosh{k }(\sinh{2k} - 2k)} 
    \Big{(} k (z - 2)\cosh{k z}
    + \sinh{k (2 - z) } - \sinh{k z} + k z \cosh{k (2 - z)} \Big{)}, \\
    \hat{w}_{\alpha} =& \frac{4\pi i k_{\alpha} \cosh{(1 - h)k}}{\cosh{k } 
    (\sinh{2k} - 2k)} 
    \left(\left(z - 2\right)\sinh{k z}- z \sinh{k (2  - z)} \right).
\end{align}
\end{subequations}
Similarly for a Stokeslet, \eqref{eq:wsource2}) transforms to give 
\begin{align}
    \hat{w}_j^k \Big{|}_{z = 0} &= 4\pi i \left( \delta_{j 3}\delta_{k \alpha} + \delta_{k 3}\delta_{j \alpha} \right) \frac{k_{\alpha}}{k}\frac{\partial}{\partial k}\left( \frac{\cosh{k(1 - h)}}{\cosh{k}} \right), \nonumber \\
    \frac{\partial \hat{w}^k_{\alpha}}{\partial z} \Big{|}_{z = 1} &= 4\pi i \, \delta_{k 3}\frac{k_{\alpha}}{k}\frac{\partial}{\partial k}\left( \frac{k \sinh{hk}}{\cosh{k}} \right), \nonumber \\
    \hat{w}_3^k \Big{|}_{z = 1} &= 4\pi \, \delta_{k3}  \Bigg{(} \frac{\partial}{\partial k}\left( \frac{\sinh{hk}}{\cosh{k}} \right) - \frac{\sinh{hk}}{k \cosh{k}} \Bigg{)}.
\end{align}
with corresponding full solution for $\hat{w}^k_j$
\begin{subequations}
\begin{align}
\hat{w}^3_3 &= \frac{8\pi}{k\cosh^2{k}\left( \sinh{2k} - 2k\right)}\Big{(}k^2\sinh{hk}\sinh{k z} 
+ hk^2z \cosh^2{k}\sinh{hk}\sinh{k z} \nonumber \\
&\hphantom{{}=\frac{8\pi}{k\cosh^2{k}\left( \sinh{2k} - 2k\right)}}+ hk^2\cosh{k}\sinh{k z}\sinh{k(1-h)} 
+ k^2z \cosh{k}\sinh{hk}\sinh{k(1-z)} \nonumber \\
&\hphantom{{}=\frac{8\pi}{k\cosh^2{k}\left( \sinh{2k} - 2k\right)}}- hk^2 z \cosh^3{k}\cosh{k(1-h-z)} 
+ hk\cosh^2{k}\cosh{hk}\sinh{k z} \nonumber \\
&\hphantom{{}=\frac{8\pi}{k\cosh^2{k}\left( \sinh{2k} - 2k\right)}}+ k z \cosh^2{k}\sinh{hk}\cosh{k z} 
- 2k\cosh{k}\sinh{k}\sinh{hk}\sinh{k z} \nonumber \\
&\hphantom{{}=\frac{8\pi}{k\cosh^2{k}\left( \sinh{2k} - 2k\right)}}- \cosh^2{k}\sinh{hk}\sinh{k z}\Big{)},
\end{align}
\begin{align}
\hat{w}^3_{\alpha} &= \frac{8\pi i k_{\alpha}}{k \cosh^2{k}\left( \sinh{2k} - 2k \right)}\Big{(} z \cosh^2{k}\sinh{hk}\sinh{k z} 
+ k\sinh{hk}\cosh{k z}\nonumber \\
&\hphantom{{}=\frac{8\pi i k_{\alpha}}{k \cosh^2{k}\left( \sinh{2k} - 2k \right)}}+ hk z \cosh^2{k}\sinh{hk}\cosh{k z} 
+ hk z \cosh^3{k}\sinh{k(1 - h - z)} \nonumber \\
&\hphantom{{}=\frac{8\pi i k_{\alpha}}{k \cosh^2{k}\left( \sinh{2k} - 2k \right)}}- \cosh{k}\sinh{hk}\sinh{k(1+z)} 
- h\cosh^2{k}\sinh{k}\sinh{k(1-h-z)} \nonumber \\
&\hphantom{{}=\frac{8\pi i k_{\alpha}}{k \cosh^2{k}\left( \sinh{2k} - 2k \right)}}+ hk\cosh{k}\cosh{k z}\sinh{k(1-h)} 
- k z \cosh{k}\sinh{hk}\cosh{k(1-z)}\Big{)}, \\
\hat{w}^{\alpha}_3 &= \frac{4\pi ik_{\alpha}}{k(\sinh{2k} - 2k)}\left( \frac{\partial}{\partial k}\left( \frac{\cosh{k(1-h)}}{\cosh{k}} \right) \right)
\Big{(} k(z-2)\cosh{k z} + k z \cosh{k(2-z)} \nonumber \\
&\hphantom{{}=\frac{4\pi ik_{\alpha}}{k(\sinh{2k} - 2k)}\left( \frac{\partial}{\partial k}\left( \frac{\cosh{k(1-h)}}{\cosh{k}} \right) \right)} + \sinh{k(2-z)} - \sinh{k z}\Big{)}, \\
\hat{w}^{\alpha}_{\beta} &= \frac{4\pi k_{\alpha}k_{\beta}}{k(\sinh{2k} - 2k)}\left( \frac{\partial}{\partial k}\left( \frac{\cosh{k(1-h)}}{\cosh{k}} \right) \right) 
\left( z \sinh{k(2-z)} - (z - 2)\sinh{k z} \right). \label{eq:fullverticalstokeslet}
\end{align}
\end{subequations}
Rewriting the inverse Fourier transform in terms of Hankel transforms, we obtain for the source
\begin{equation}
    w_3 = \frac{1}{2\pi} \mathcal{H}_0 \left( \hat{w}_3 \right), \quad w_{\alpha} 
    = \frac{i x_{\alpha}}{2\pi\rho} \mathcal{H}_1 \left( \frac{k \hat{w}_{\alpha}}{k_{\alpha}} \right), 
\end{equation}
and for the Stokeslet
\begin{subequations}
\begin{align}
    w_3^3 &= \frac{1}{2\pi}\mathcal{H}_0\left(\hat{w}^3_3\right), \quad w_{\alpha}^3 = \frac{i x_{\alpha}}{2\pi \rho}\mathcal{H}_1\left( \frac{k}{k_{\alpha}}\hat{w}^3_{\alpha} \right), 
    \quad 
    w_{3}^{\alpha} = \frac{i x_{\alpha}}{2\pi \rho}\mathcal{H}_1\left( \frac{k}{k_{\alpha}}\hat{w}^{\alpha}_{3} \right), \label{eq:hankelw3alpha} \\
w_{\beta}^{\alpha} &= \frac{1}{2\pi}\left( \frac{\delta_{\alpha \beta}}{\rho} - 2\frac{x_{\alpha} x_{\beta}}{\rho^3} \right) \mathcal{H}_1\left( \frac{k}{k_{\alpha} k_{\beta} }\hat{w}^{\alpha}_{\beta} \right)+ \frac{x_{\alpha}x_{\beta}}{2\pi \rho^2}  \mathcal{H}_0\left( \frac{k^2}{k_{\alpha} k_{\beta} }\hat{w}^{\alpha}_{\beta} \right), \label{eq:hankelwbetaalpha}
\end{align}
\end{subequations}
where $\alpha \in [1,2]$ and $\mathcal{H}_{i}$ is the Hankel transform of order $i$. Similar 
integral expressions in terms of Hankel transforms can be constructed for other Stokes singularities 
(see Appendix \ref{rotletappendix} for the rotlet, \ref{stressletappendix} for the stresslet, \ref{rotletdipoleappendix} for the rotlet dipole, and \ref{sourcedipoleappendix} for the source dipole). 

\begin{figure*}[t]
\centering\includegraphics[trim={0 0cm 0 0cm}, clip, width=0.98\columnwidth]{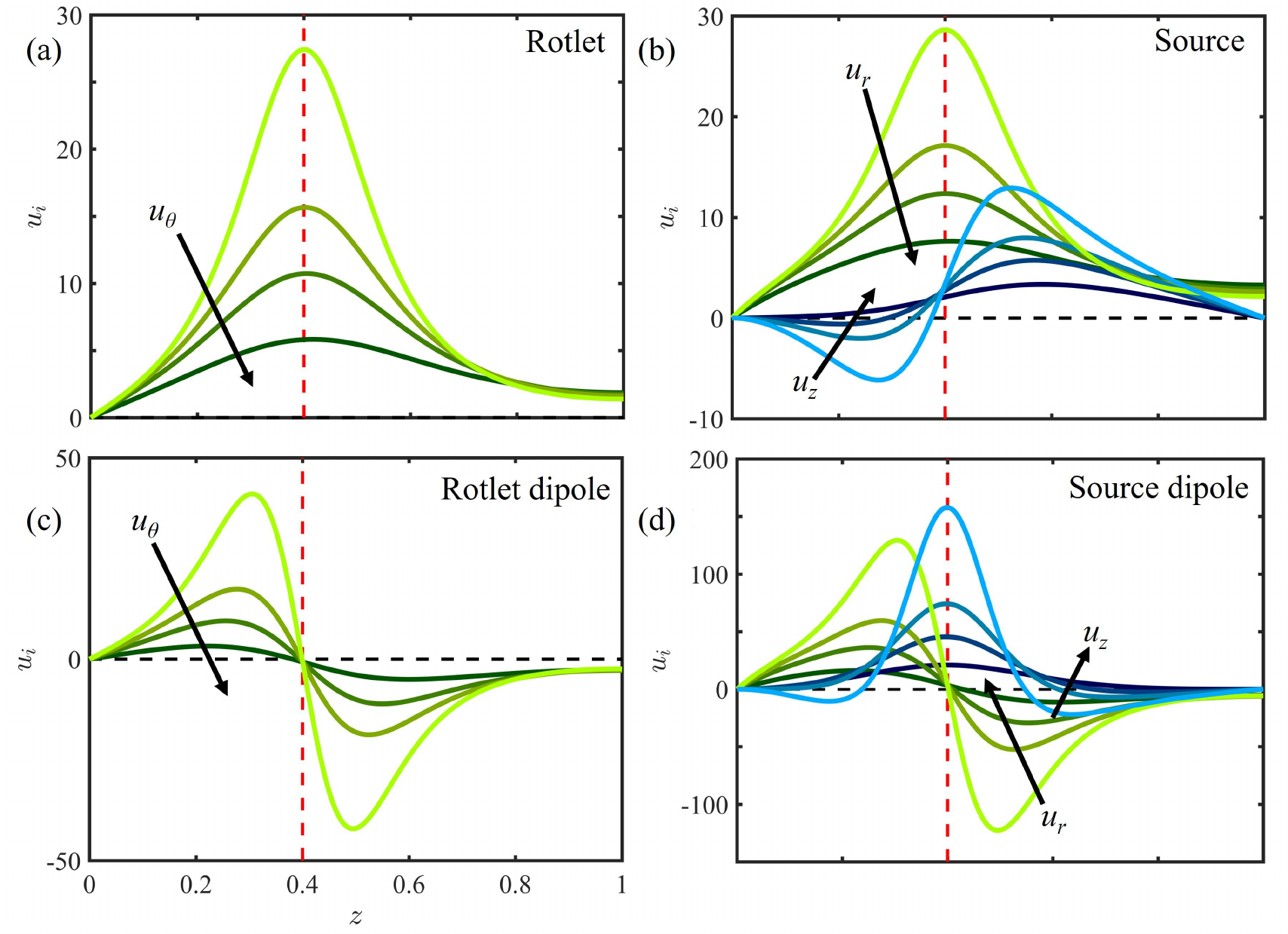}
\caption{The near field velocity $u_i$ produced by a number of singularities at 
$h = 0.4$ as a function of $z$ for a range of $x \in \{ 0.19, \, 0.25, \, 0.3, \, 0.4 \}$, $y = 0$, with darker colours denoting larger $x$. (a) Rotlet, $i = \theta$ (green curves) (b) Source, $i = r$ (green) or $i = z$ (blue) (c) Rotlet dipole, $i = \theta$ (green) (d) Source dipole, $i = r$ (green) or $i = z$ (blue). Note that here $(r, \theta)$ are the polar coordinates for the horizontal plane i.e. $x = r \cos{\theta}$ and $y = r \sin{\theta}$.
}
\label{fig3}
\end{figure*}

To illustrate the nature of these exact solutions, Fig.~\ref{fig3} plots various components of the 
fluid velocity field induced by four of the main 
singularities, the rotlet, source, rotlet dipole and 
source dipole, as a function of vertical height $z$ 
for a range of horizontal radial distances away from 
the singularities, in each case located at $h=0.4$. 

For the swirling component of the flow due to a rotlet, 
Fig.~\ref{fig3}(a) illustrates clearly how the 
boundary conditions of no slip and no stress are 
satisfied, and the incipient divergence as the $x$ location approaches that of the singularity.  
For the 
source in Fig.~\ref{fig3}(b) the horizontal velocity $u_x$
displays an increasing maximum as the observation point $x$ approaches 
the singularity location, while the vertical velocity component $u_z$ 
has a positive divergence for $z\to h^+$ and a
negative divergence as $z\to h^-$ as expected for a source, while vanishing at the top and bottom boundaries, as required by \eqref{final_bcs}.
Both the rotlet dipole in Fig.~\ref{fig3}(c) and the
source dipole in Fig.~\ref{fig3}(d) appear as derivatives
of their corresponding monopoles.

\begin{figure}[t]
\centering\includegraphics[trim={0 0cm 0 0cm}, clip, width=0.5\columnwidth]{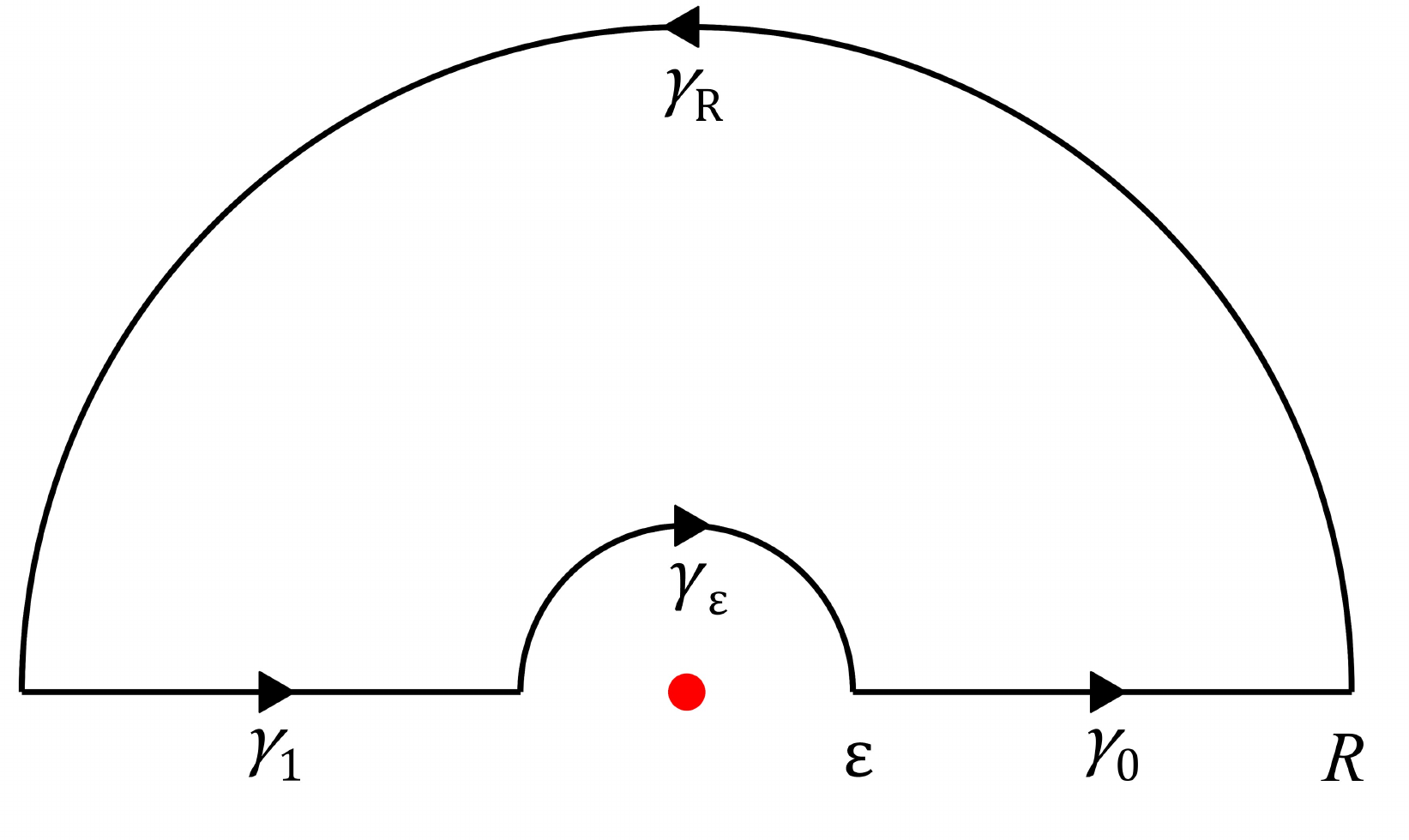}
\caption{The notched semicircular contour $\gamma$.}
\label{fig4}
\end{figure}

\section{Far-field Solutions} \label{farfieldsolution}

It is difficult to find the far-field ($\rho \gg 1$) behaviour of these solutions 
when they are expressed as exact solutions in integral form as Hankel transforms. 
Following the approach of Liron and Mochon \cite{Liron76}, we may utilise a contour integration to 
express the exact solutions in series form. Given an even function $f(z)$ decaying exponentially to zero 
on the real axis as $z = x \rightarrow \pm \infty$, consider the contour integral $\oint_{\gamma} F$ where 
$F = z^{i + 1}f(z) H^1_{i}(\rho z), H^1_{i} = J_{i} + i Y_{i}$ with $i \in [0 \, , \, 1]$ is a 
Hankel function of the 1st kind and $\gamma = \gamma_0 + \gamma_1 + \gamma_R + \gamma_{\epsilon}$ is a 
notched semicircular contour centered at the origin (Fig.~\ref{fig4}). From Watson \cite{Watson22}, 
$\int_{\gamma_{R}}F \rightarrow 0$ as $R \rightarrow \infty$. Hence, applying the residue theorem 
in the limit as $R \rightarrow \infty$ and $\epsilon \rightarrow 0$ yields
\begin{equation}
    \int^{\infty}_{0} \lambda^{i+1} d\lambda \, J_{i}(\lambda\rho) f = - \frac{1}{2}\int_{\gamma_{\epsilon}} F 
    + \pi i \sum \mbox{Residues of singularities of F in } \gamma. \label{eq:residuetheorem}
\end{equation}

Using this method, the repeated reflection solutions $v_j$ for all four primary Stokes singularities can be directly expressed in series form. For a source, $v_j$ becomes
\begin{subequations}
\begin{align}
    v_3 &= -2\pi \sum_{n = 1, 3, 5, \ldots}^{\infty}\!\!\!\!\!\! n 
    \sin{\left( \frac{n \pi h }{2} \right)}\cos{\left( \frac{n \pi z}{2} \right)}
    K_0\left( \frac{n \pi \rho}{2} \right), \\
    v_{\alpha} &= \frac{2\pi x_{\alpha}}{\rho}\sum_{n = 1, 3, 5, \ldots }^{\infty}\!\!\!\!\!\! n 
    \sin{\left(\frac{n \pi h}{2} \right)}\sin{\left( \frac{n \pi z}{2}\right)}
    K_1 \left( \frac{n \pi \rho}{2} \right).  
\end{align}
\end{subequations}

\begin{figure*}
\centering\includegraphics[trim={0 0cm 0 0cm}, clip, width=1.0\textwidth]{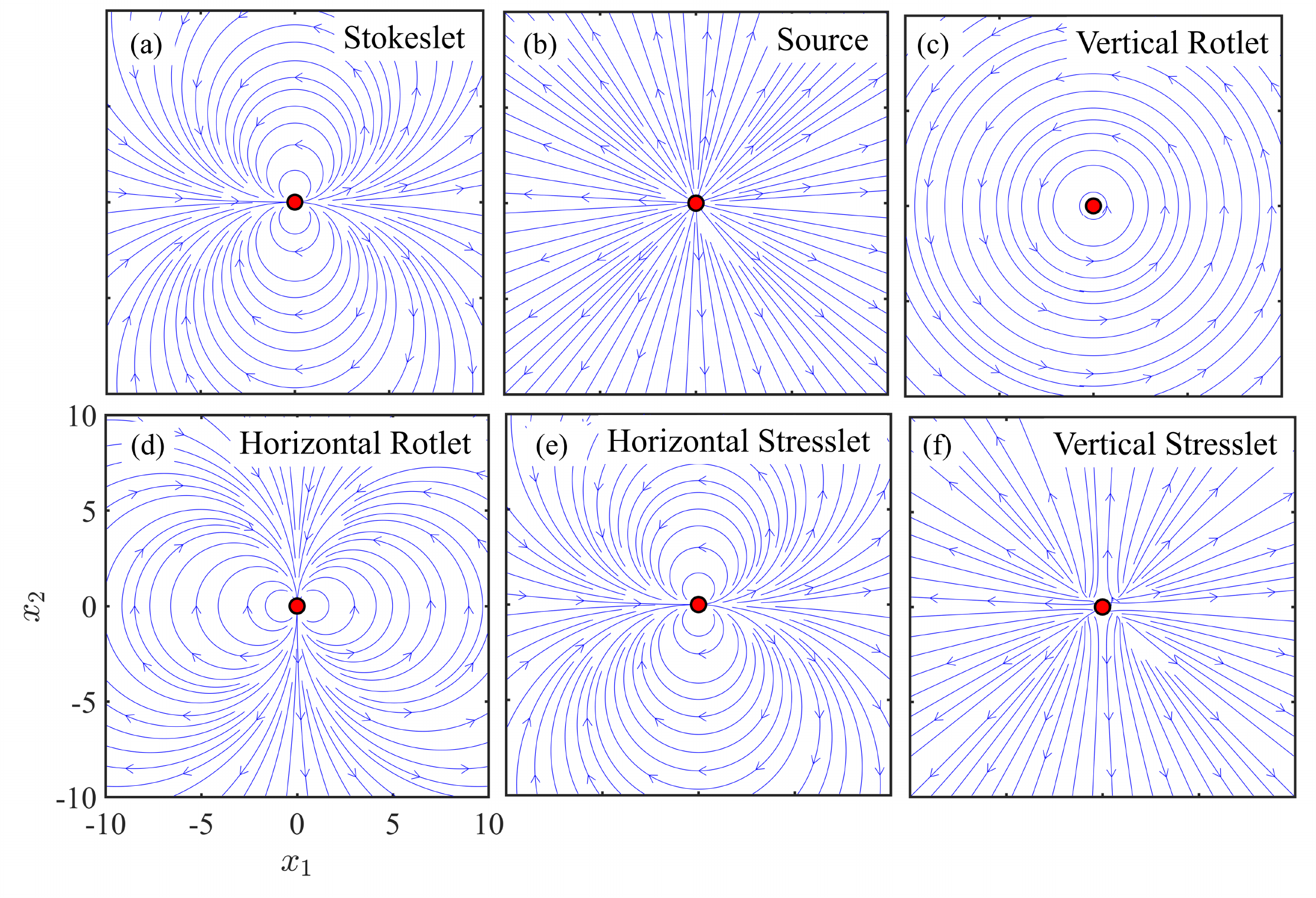}
\caption{Streamlines in the $z = 1$ plane for the flows generated by Stokes singularities in 
the far field thin-film limit ($\rho \gg H$): (a) Stokeslet orientated in the positive $x$ direction, (b) source, (c) and (d) Rotlet orientated in the $z$ and $x$ directions, respectively, (e) and (f) 
Stresslet $u^{k,l}$ with $k = 1, l = 3$ and $k = l = 1$, respectively. As streamlines in (f)
depend on $h$, we have set $h = 1/2$.}
\label{fig5}
\end{figure*}

Note that for all four singularities, the dominant term in the far-field expansion ($\rho \gg 1$) of the repeated reflection solution $v_j$ comes from the $n = 0$ terms and decays like exp$(-\pi\rho/2)$. Similarly, the integral expressions for the auxiliary solution $w_j$ can be expressed in series form to obtain series expansions for the full flow field $u_j$. For a source, the corresponding complex function F has in $\gamma$ poles of order 1 at $z = \pi i(n + 1/2)$ where $n \in \mathbb{Z}^{\geq}$ and poles of order 1 at $z = z_0/2$ where $z_0$ satisfies $\sinh{z_0} = z_0$. Since $\int_{\gamma_{\epsilon}}$ vanishes as $\epsilon \rightarrow 0$, when $j = k = l = 3$ (\ref{eq:residuetheorem}) simplifies to become
\begin{align}
    w_3 =& 2\pi (1 - z)\!\! \sum_{n = 1, 3, 5, \ldots}^{\infty}\!\!\!\!\!\! n \sin{\left( \frac{n \pi h}{2} \right)} \cos{\left( \frac{n \pi z}{2} \right)}K_0 \left( \frac{n \pi\rho}{2} \right)
    + \mathcal{O}\left( \frac{e^{-\rho y_1 / 2}}{\sqrt{\rho}} \right) \\
    u_3 =& v_3 + w_3 \nonumber \\
    =& -2\pi z \sum_{n = 1, 3, 5, \ldots}^{\infty} \!\!\!\!\!\! n \sin{\left( \frac{n \pi h}{2} \right)} \cos{\left( \frac{n \pi z}{2} \right)}K_0 \left( \frac{n \pi\rho}{2} \right).
\end{align}

The first term dominates in the far-field, so 
\begin{equation}
    u_3 \simeq
    - \frac{2\pi z}{\sqrt{\rho}} \cos{\left( \frac{\pi z}{2} \right)}  \sin{ \left( \frac{\pi h}{2} \right)} e^{-\rho \pi / 2} + \mathcal{O}\left( \frac{e^{-\rho \pi / 2}}{\rho^{3/2}} \right), \label{eq:verticalsource}
\end{equation}
namely an exponential radial decay with $z$ dependence  $z\cos{(\pi z / 2)}$, vanishing at both surfaces. Furthermore, when $j = \alpha \in [1,2]$, the leading order contribution in the far field arises from $\gamma_{\epsilon}$, namely
\begin{equation}
    u_{\alpha} = z\left( 2 - z \right)\left[ \frac{3 x_{\alpha}}{\rho^2} \right], \label{eq:sourcefarfieldform}
\end{equation}
noting that the contribution from the poles at $z = \pi i (n + 1/2)$ in $w_{\alpha}$ cancels out with $v_{\alpha}$. 
Similarly for a Stokeslet, $F$ has poles of order 2 at $z = \pi i(n + 1/2)$ where $n \in \mathbb{Z}^{\geq}$ and poles of order 1 at $z = z_0/2$ where $z_0$ satisfies $\sinh{z_0} = z_0$. When $j = k = 3$, since $\int_{\gamma_{\epsilon}}$ vanishes as $\epsilon \rightarrow 0$, \eqref{eq:residuetheorem} simplifies to 
\begin{align}
 w^3_3 &= -\sum_{n = 1, 3, 5, \ldots}^{\infty}\sin{\Bigg{(} \frac{n \pi h}{2}\Bigg{)}}\sin{\Bigg{(}\frac{n \pi z}{2}\Bigg{)}} 
 \Bigg{(} 8K_0\Bigg{(}\frac{n \pi\rho}{2}\Bigg{)} - 2n \pi\rho K_1\Bigg{(} \frac{n \pi\rho}{2} \Bigg{)}\Bigg{)} \nonumber \\
&+ \sum_{z_0 \in \mathbb{H} \colon z_0 = \sinh{z_0}} \frac{i z_0}{8(\cosh{z_0}-1)}
 \Big{(} \hat{w}^3_3 (\sinh{2k} - 2k) \Big{)}\Big{|}_{k = z_0/2} H_0^1\left(\frac{\rho z_0}{2}\right), \nonumber \\
 u_3^3 &= v_3^3 + w^3_3 = \sum_{z_0 \in \mathbb{H} \colon z_0 = \sinh{z_0}} \frac{i z_0}{8(\cosh{z_0} - 1)}
 \times \Big{(} \hat{w}^3_3 (\sinh{2k} - 2k) \Big{)}\Big{|}_{k = z_0/2} H_0^1\left(\frac{\rho z_0}{2}\right),
\end{align}
noting that the contribution from the poles of order $2$ in $w^3_3$ cancels out with $v^3_3$. The leading far-field behavior is
\begin{equation}
    u^3_3 = \mathcal{O}\left( \frac{e^{-\rho y_1/2}}{\rho^{1/2}}  \right), 
\end{equation}
where $y_1 = 7.498\ldots$ is the imaginary part of the first non-zero root to $\sinh{z_0} = z_0$ in the first quadrant. Similarly for $j = \alpha, \, k = 3$ and  $k = \alpha, \, j = 3$ where $\alpha \in [1 \, , \, 2]$, the leading order 
far-field contribution is
\begin{equation}
u^{3}_{\alpha}, \, u^{\alpha}_3  = \mathcal{O}\left( \frac{x_{\alpha} \, e^{-\rho y_1/2}}{\rho^{3/2}} \right). \label{eq:horizontalcomponentverticalStokeslet}
\end{equation}
When $j = \beta$ and $k = \alpha$ where $\alpha \, , \beta \in [1 \, , \, 2]$, the leading order contribution in the far-field arises from $\gamma_{\epsilon}$, 
\begin{align}
    u^{\alpha}_{\beta} &= z(2-z)\left[ -\frac{3h(2-h)}{\rho^2}\left( \delta_{\alpha\beta} - \frac{2x_{\alpha}x_{\beta}}{\rho^2} \right) \right]. \label{eq:stokesfarfieldform}
\end{align}
Similar far-field approximations can be found for the other Stokes singularities 
(Appendix \ref{rotletappendix}, rotlet; \ref{stressletappendix}, stresslet; \ref{rotletdipoleappendix}, rotlet dipole; \ref{sourcedipoleappendix} source dipole).

Figure.~\ref{fig5} plots streamlines of these far field flows
in the horizontal plane $z = 1$. In Fig.~\ref{fig5}(a), 
a Stokeslet orientated in the $x$ direction generates a flow with a recirculating flow pattern of two 
loops decaying radially like $1/\rho^2$, namely a two dimensional source dipole 
(recalling that the source flow $u_{s} = x_i / \rho^2$ leads to the source dipole flow 
$u_{sd} = \delta_{i j}/\rho^2 - 2x_ix_j/\rho^4$). Confinement has fundamentally affected the 
unidirectionality of the flow by inducing recirculation in the $y$ direction. This  is a 
feature of the family of Stokes singularities that are derivatives of the Stokeslet, 
with higher order singularities having more recirculation loops.  For example, a Stokes dipole has four loops while 
a Stokes quadrupole has six.  In contrast, the spherical symmetry of a three dimensional source ensures that the new flow 
is still a source (Fig.~\ref{fig5}(b)). 
Derivatives of the source, such as the source dipole, are also unchanged by confinement, and  
since the vertically orientated rotlet is independent of $z$, its streamlines are also unchanged, as seen in Fig.~\ref{fig5}(c). Confinement breaks the symmetries of the horizontal 
rotlet and stresslet, leading to flows with the character a two dimensional source dipole for both a horizontally 
orientated rotlet (Fig.~\ref{fig5}(d)) and a vertical stresslet ($j = 1$, $k = 3$, Fig.~\ref{fig5}(e)) 
and a two dimensional source for a horizontal stresslet ($j = k = 3$, Fig.~\ref{fig5}(f)), respectively. 

\section{Leading Order Far Field Flow} \label{discussion}
\label{discuss}
Examining the cases given above in \S\ref{farfieldsolution} and in Appendices 
(\ref{rotletappendix})-(\ref{sourcedipoleappendix}), we note that for the four primary Stokes 
singularities, the leading order far-field flow is separable in $z$ (formally considering the 
limit where $h, \, H, z$ are fixed while $\rho$ is large). If the flow does not decay 
exponentially radially, the it has $z$ dependence of the form $z (1 - z/2)$. Otherwise, 
the flow decays exponentially either as $\exp{(-\rho\pi/2)}$, arising from a $K_1(\rho\pi/2)$ 
term with corresponding $z$ dependence either $\sin{\pi z / 2}$ for horizontal flow or 
$z \cos{\pi z / 2}$ for vertical flow, or $\exp{(-\rho y_1/2)}$ where $y_1 \approx 7.498$ 
is the imaginary part of the first non-zero root to $\sinh{z_0} = z_0$ in the upper half 
plane. All higher order Stokes singularities can be expressed as derivatives of these 
four primary Stokes singularities. These singularities must also either have leading 
order $z$ dependence $z (1 - z/2)$ or decay exponentially like $\exp{(-\rho\pi/2)}$ 
or $\exp{(-\rho y_1/2)}$. This means that the leading order far field contribution 
to the flow from these singularities can be obtained directly by differentiating the 
far field flows for the primary Stokes singularities, namely the full exact solutions 
which quickly become very complicated do not need to be derived. For example, 
differentiating (\ref{eq:stokesfarfieldform}) once, (\ref{eq:stokesfarfieldform}) 
twice and (\ref{eq:sourcefarfieldform}) once recovers the far field flows for a Stokes 
dipole, a Stokes quadrupole and a source dipole respectively given in \cite{Mathijssen16}, noting 
a sign error there in the expression given for a Stokes quadrupole (their equation (B8)), namely
\begin{align}
\left[u^j_{D} \right]_i &= \frac{6}{\rho^4}\left( x_{i} +2 x_{j} \delta_{ij} - \frac{4 x_i x_j^2}{\rho^2} \right) h z \left(2- h \right)\left(2-z\right), \\
\left[u^j_{Q} \right]_i &= \frac{18}{\rho^4}\left( \delta_{ij} - \frac{4 x_{i}x_{j}}{\rho^2} - \frac{4 x_{j}^2 \delta_{i j}}{\rho^2} + \frac{8 x_i x_j^3}{\rho^4} \right) \times h z \left(2- h \right)\left(2-z \right), \\
\left[ u_S^j \right]_i &= \frac{6z}{\rho^2}\left( \delta_{ij} - \frac{2 x_i x_j}{\rho^2} \right)\left( 1 - \frac{z}{2} \right). \label{eq:maintextsourcedipole}
\end{align}
As a consistency check, (\ref{eq:maintextsourcedipole}) does indeed reproduce what was derived from first principles in Appendix \ref{sourcedipoleappendix}. Hence, for an arbitrary body whose free-space locomotion can be captured by a expansion in terms of Stokes singularities, the far field flow field is separable in $z$ with either $z$ dependence of the form $z(1 - z/2)$ or the flow decays radially exponentially. The fluid velocity field $\boldsymbol{u}$ can thus be factorised as $\boldsymbol{u} = f(z)\boldsymbol{U}(\boldsymbol{x_h})$ where $\boldsymbol{x_h} = (x_1,x_2)$ and $f(z)$ is normalised so that $\left(\int^H_0 f \, dz \right) = 1$ (typically f is either $3z(1 - z/2)$ or $\pi \sin{(\pi z / 2 )}/2$). The 3D Stokes equations for $\boldsymbol{u}$ reduces to a Brinkman-like equation for the vertically averaged fluid velocity $\boldsymbol{U}$
\begin{equation}
    \mu \left( \nabla^2 - \kappa^2 \right) \boldsymbol{U} = \nabla p,
\end{equation}
with corresponding incompressibility condition
$\boldsymbol{\nabla \cdot U} = 0$,
where $\kappa = \left(\partial f/\partial z |_{z = 0} \right)^{1/2}$ plays the role of the inverse Debye screening length in screened electrostatics. We have thus reduced a 3D 
system to a 2D one that can be solved by transforming to an appropriate coordinate 
system that simplifies the boundary conditions. This method is equally applicable 
in the setup of Liron and Mochon \cite{Liron76}, namely a microfluidic environment 
between two horizontal rigid boundaries, where the corresponding far field $z$ 
dependence for an non radially exponentially decaying flow is $z (1 - z)$.   

\section{Self-consistency Check}

A key assumption made above was that the combination of surface tension and gravitational effects restricts vertical deformation of the interface and hence $H$ can be assumed constant. As a self-consistency check, using (\ref{eq:verticalsource}), the leading order contribution to the stress $\sigma_{x_3 x_3}$ in the far field at the upper free surface boundary $x_3 = H$ that a source of strength $\lambda_{S}$ (namely generating a flow $u_i = \lambda_{S} x_i / r^3$) at $(x_1, \, x_2, \, x_3) = (0,0,s)$ produces is
\begin{align}
\sigma_{x_3 x_3} &= 2\mu \frac{\partial u_3}{\partial x_3} = \frac{\mu \pi^2}{H^3}\sin{\left( \frac{\pi s}{2H} \right)}K_0\left( \frac{\rho \pi}{2H} \right) \nonumber \\
&\leq \frac{\mu \pi^2}{H^3}K_0\left( \frac{\rho \pi}{2H} \right) \approx \frac{\mu \pi^2}{\rho^{1/2}H^{5/2}}e^{-\rho \pi / 2H},
\end{align}
when $\rho \gg 2H/\pi$. Here, we have utilized the asymptotic large argument expansion for $K_{\alpha}$ \cite{Abramowitz70} together with the fact that $|\sin(\pi s / 2H)| \leq 1 \, \forall \, s \in [0, \, H]$. Hence, a measure $M_{s}$ of the relative strength of the stresses at the free surface arising from the flow generated by the singularity that seek to deform this surface to the gravitational forces restricting vertical deformation is
\begin{equation}
M_{s} = \frac{\mu \pi^2 \lambda_{S}}{\rho_{w}\rho^{1/2}g H^{5/2}\Delta H}e^{-\rho \pi / 2H}.
\end{equation}
Writing the strength of the source $\lambda_{S}$ as $\lambda_{S} = U_{S}H^2$, $U_s$ scales with the typical velocities of flows in a Petri dish, namely $U_S \sim \SI{2}{\milli\metre \second^{-1}}$. Hence, setting $\mu = \SI{1}{\milli\pascal \, \second^{-1}}, \, H = \SI{5}{\milli\metre}, \, \Delta H = \SI{0.1}{\milli\metre}, \, \rho = \SI{1}{\centi\metre}$ we find
    $M_s \simeq 1.2 \times 10^{-4} \ll 1$, so
$M_s$ is indeed small and thus the flat surface approximation is consistent for a source.

\section{Case Study: Hydrodynamic Bound States}
\label{boundstates}

An instructive application of the results of this paper is exploring the notion of ``hydrodynamic bound states".  First discovered 
by Drescher, {\it et al.} in 2009 using the green alga
{\it Volvox} \cite{Drescher09}, these are dynamical states exhibited by pairs of
spherical chiral microswimmers near a surface.  {\it Volvox}
colonies have radius $R \sim\!\! 250\,\mu$m, with $\sim\!\! 10^3$
biflagellated somatic cells beating on their surface. This beating is primarily
in the posterior-anterior direction, but has a modest orthogonal
component that leads to spinning motion about the AP axis.
While the organisms are slightly denser than the fluid surrounding
them, the flagellar beating allows them to swim upwards against
gravity.  When a suspension of {\it Volvox} was placed in a
glass-topped chamber, the colonies naturally swam upwards due
to their bottom-heaviness, which aligned their AP axis with
gravity.  Pairs of colonies at 
the chamber top were found to move together
while they continued to spin, eventually touching and 
orbiting about each other.  

\begin{figure}[t]
\centering\includegraphics[trim={0 0cm 0 0cm}, clip, width=0.7\columnwidth]{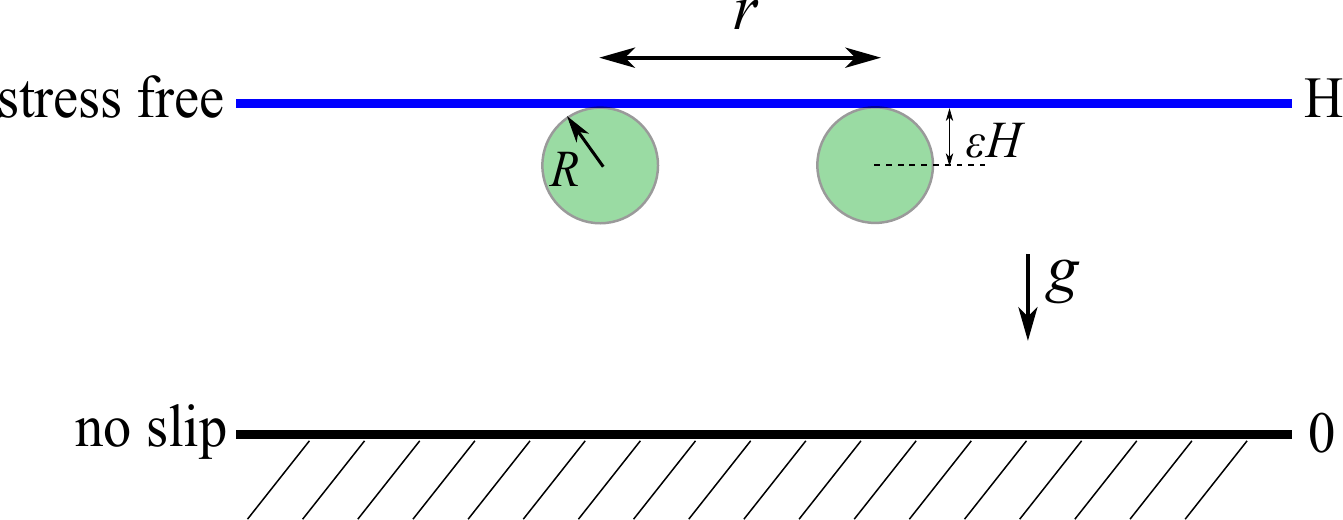}
\caption{Geometry of hydrodynamic bound states. Two spherical, negatively buoyant 
microswimmers of radius $R$ just below an upper surface, a horizontal distance $r$ apart.}
\label{fig6}
\end{figure}

As shown schematically in Fig.~\ref{fig6}, once the colonies 
have ascended as high as possible, their centers are a distance 
$R = \epsilon H$ (with $\epsilon \ll 1$) below 
the upper no-slip surface.  Due to their positive density offset
relative to the surrounding ambient water, they are acted on by a
downward gravitational force.  Viewed from afar, each colony can be
considered as a point force acting on a fluid: the resultant flow 
field is that of a downward-pointing Stokeslet
of magnitude $F=(4\pi/3)R^3\Delta\rho g$ associated with the
gravitational force.
This geometry\textemdash two nearby Stokeslets directed away from a 
no-slip wall\textemdash is exactly that envisioned by Squires
\cite{Squires} in his analysis of surface-mediated interactions, who showed that the mutual advection of
those Stokeslets toward each other is described by the dynamics of
their separation $r$ in the form
\begin{equation}
    \dot{r} = -\frac{3F}{\pi\mu R}\frac{rR^4}{(r^2 + 4 R^2)^{5/2}},
    \label{infalling_noslip}
\end{equation}
expressed in a way that identifies the characteristic 
speed $F/\mu R$. Tracking of  
{\it Volvox} pairs showed precise quantitative agreement with this
result \cite{Drescher09}.  While it was not clear {\it a priori} that the 
Stokeslet approximation was valid
over the large range of inter-colony separations explored, 
direct measurements of the flow fields around freely swimming
colonies \cite{Drescher10} showed that the
Stokeslet does indeed dominate all higher-order singularities 
beyond a few radii from the colony center.

This general phenomenon has been rediscovered several times:
in suspensions of the fast-moving bacterium {\it Thiovulum majus} 
\cite{Petroff15}, of the magnetotactic bacterium 
{\it Magnetotacticum magneticum} \cite{Pierce18}, and of 
starfish embryos \cite{Tan22}.  In the latter case, the
pairwise bound states occur at the air-water interface,
which can be taken to be a stress-free boundary.
In that case, and for an infinitely deep fluid, the image system
for each Stokeslet is simply an opposite Stokeslet above the
air-water interface - singularity $2$ in Fig.~\ref{fig2}.
Thus, the lateral flow at ($x_1,0,x_3$) due to a downward Stokeslet at the origin is
\begin{equation}
    u=\frac{F}{8\pi\mu}\Bigg\{\frac{x_1(x_3-(1-\epsilon) H)}{[x_1^2+(x_3-(1-\epsilon)H)^2]^{3/2}}
    -\frac{x_1(x_3-(1+\epsilon) H)}{[x_1^2+(x_3-(1+\epsilon)H)^2]^{3/2}}\Bigg\}.
    \label{george1}
\end{equation}
If we evaluate this flow at the Stokeslet location $x_3=(1-\epsilon) H$, 
and multiply by a factor of $2$ we obtain the dynamics of the
particle separation $r$ in a 
form similar to the no-slip result \eqref{infalling_noslip},
but with a different power law exponent in the denominator,
\begin{equation}
    \dot{r} = -\frac{F}{2\pi\mu R}\frac{r R^2}{(r^2 + 4 R^2)^{3/2}},
    \label{infalling_nostress}
\end{equation}
where $R = \epsilon H$. In each of \eqref{infalling_noslip} and \eqref{infalling_nostress}
we can identify an effective potential energy $V(r)$ such that
$\dot{r}=-dV/dr$.  
A natural question is how the result \eqref{infalling_nostress} for a stress-free surface is modified in the 
geometry of a Petri dish. The three lengths $R= \epsilon H$, $H$, 
and $r$ must be compared to determine the appropriate asymptotic regime.

\begin{figure}
\centering\includegraphics[trim={0 0cm 0 0cm}, clip, width=0.7\columnwidth]{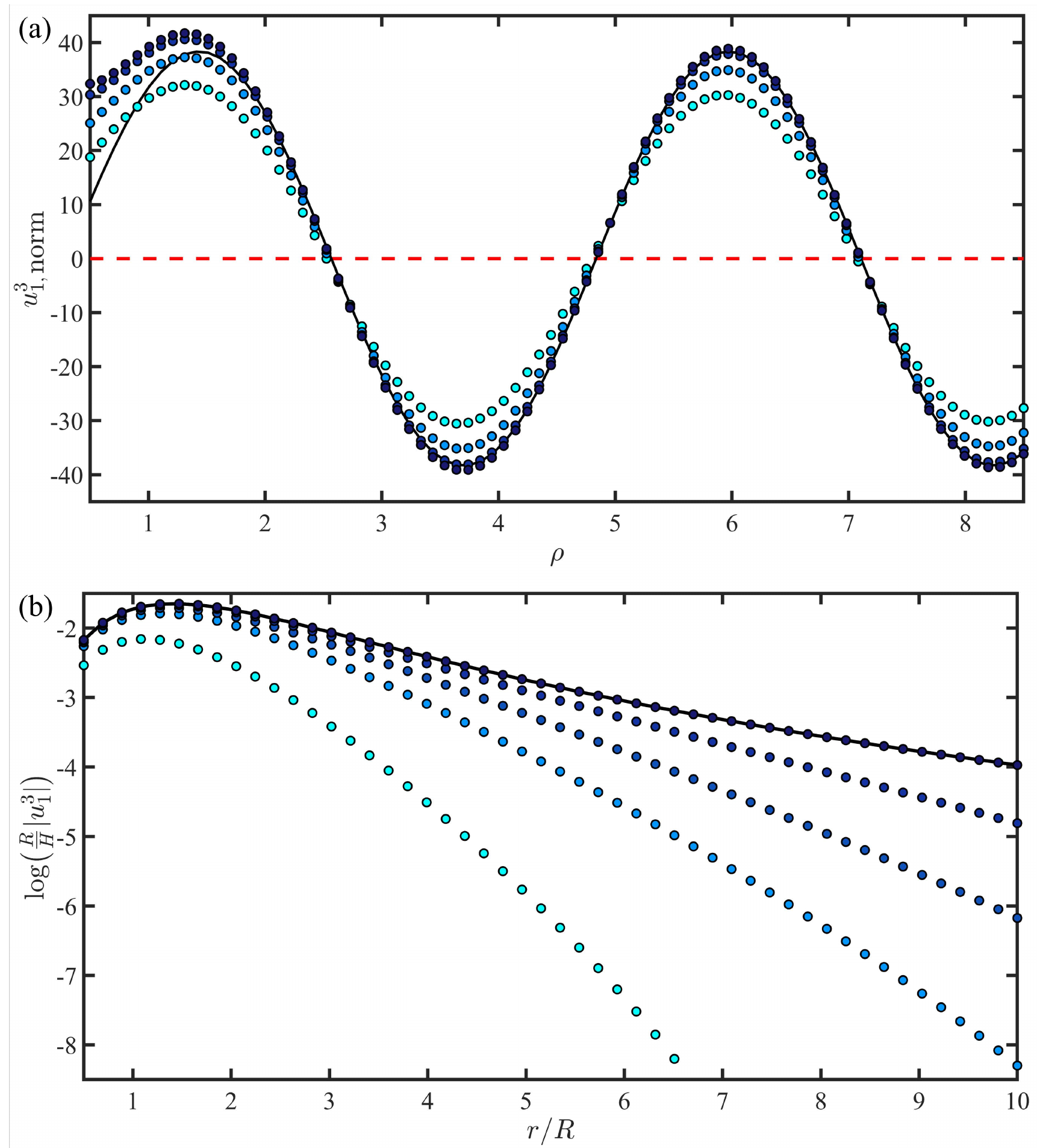}
\caption{The lateral flow leading to hydrodynamic bound states. 
(a) Numerically obtained horizontal fluid velocity 
field $u^3_{1,\mbox{norm}} = (\sqrt{\rho}/\epsilon) e^{\rho y_1/2}u^3_1$, 
normalized to highlight the asymptotical sinusoidal component, 
generated by a vertically 
orientated Stokeslet placed at $(0,0,H-R)$ and evaluated 
as a function of $\rho$ at the point $(\rho,0,h)$. Here, $R/H \in [0.15,0.1,0.05,0.01]$ 
with darker shades of blue denoting smaller values of $R/H$. The similarly scaled 
asymptotic result \eqref{eq:complexasymptoticresult} is shown as the solid line. 
(b) The velocity $u^3_1$ as a function of $r/R$. Here, $R/H \in [0.3,0.2,0.15,0.1,0.01]$ 
with darker shades of blue denoting smaller values of $R/H$. For comparison the asymptotic 
result \eqref{george1} for an infinitely deep Petri dish is 
shown as the solid black line.}
\label{fig7}
\end{figure}

The dynamics \eqref{infalling_nostress} holds for $r\ll H$ but without
restriction on the relative sizes of $r$ and $R$, except that
the impenetrability of the colonies implies that this expression is only
relevant for $r>2R$.  Of course, the
validity of the singularity approach itself will decrease for 
$r\sim R = \epsilon H$, and thus it is fair to assert that \eqref{infalling_nostress}
is {\it physically} valid for $\epsilon H \ll r \ll H$, and in particular 
$\dot{r}\sim (F/2\pi\mu R)(R/r)^2$ for $r\gg R$.
Indeed, as a consistency check, the full integral expressions do indeed simplify to \eqref{infalling_nostress} in this limit as we now show. Working in the same horizontal plane as the singularity, after some contour integration the repeated reflection solution becomes 
\begin{align}
v_{\alpha}^3 &= - x_{\alpha} \int^{\infty}_0 \lambda J_0(\lambda \rho) \frac{\cosh{\lambda (1 - 2 h)}}{\cosh{\lambda}} \, d\lambda, \nonumber \\ 
&= - 2\pi x_{\alpha} \sum_{n = 0}^{\infty} \left(n + \frac{1}{2}\right) \sin{2 \pi \epsilon \left(n + \frac{1}{2}\right)}
K_0\left(\pi \rho \left(n + \frac{1}{2} \right)\right).
\end{align}
From (\ref{eq:hankelw3alpha}) we find that for small $\rho$ the auxiliary solution $w^3_\alpha$ is
$\mathcal{O}\left(x_{\alpha} \epsilon/\rho^2 \right)$
and hence, for points with small $\epsilon$ and $\rho$, the repeated 
reflection solution dominates the auxiliary solution. Expanding in powers of $\epsilon$ we 
find
\begin{equation}
u^3_\alpha = -\frac{2 \epsilon x_{\alpha}}{\rho^3} + \frac{12 \epsilon^3 x_{\alpha}}{\rho^5} + \cdots, \label{eq:boundaryapprox}
\end{equation}
a result that agrees precisely with an expansion in $\epsilon$ and suitable 
nondimensionalization of \eqref{george1}.

The new regime of interest occurs when the separation $r$ becomes 
comparable to or larger than the Petri dish depth $H$. Given for completeness 
in Appendix \ref{VerticalStokesletAppendix}, when $r \gg H$ ($\rho \gg 1$), the 
non-dimensional flow field $u^3_{\alpha}$ decays exponentially with an unusual sinusoidal 
form 
\begin{equation}
u^3_{\alpha} = \frac{A \epsilon x_{\alpha}}{\rho^{3/2}} 
e^{-\rho y_1/2}\sin{(x_1(\rho - \rho_0)}/2), \label{eq:complexasymptoticresult}
\end{equation}
where $z_1 = x_1 + i y_1 = 2.769 + 7.498 i$ is the first root in the first 
quadrant to the equation $\sinh{z_1} = z_1$, $A = 38.340$ and 
$\rho_0 = 0.298$. Figure \ref{fig7}(a) explores this further, demonstrating how 
numerical solutions to the full flow field vary as a function of $\rho$ for a range of 
values of $h$. Darker blue dots denote larger values of $h$ i.e. the {\it Volvox} 
are closer to the free surface. For comparison, the asymptotic result 
\eqref{eq:complexasymptoticresult} is superimposed on those numerical results. 
For clarity, all velocities are normalised by $(\sqrt{\rho}/\epsilon)e^{\rho y_1/2}$ to 
highlight the sinusoidal component of the flow field. As can be seen, the asymptotic 
result is a good fit for $\rho\gtrsim 2$, improving as $\rho$ increases and as $h\to 1$.

An interesting feature of the screened interaction is that the multiplicative power
law $\rho^{-1/2}$ differs from that underlying the unscreened 
form \eqref{infalling_nostress}, which falls off as $\rho^{-2}$.  This is unlike the 
case in electrostatics, for example, where a screened Coulomb interaction in three
dimensions decays as $\sim (1/r)e^{-r/\lambda}$, where $\lambda$ is the screening length,
and the unscreened interation is $\sim 1/r$.
In the present case, the reason why we see a transition as $r$ increases is that for 
small $r$ the first reflection from the repeated reflection solution dominates, 
but as $r$ increases the auxiliary solution generates terms that cancel out the repeated reflection solution, thus leaving lower order terms in the auxiliary solution to dominate, giving rise to an exponential decay. 

Figure \ref{fig7}(b) shows in a semilogarithmic plot the lateral fluid velocity $u_1^3$
as a function of the dimensionless radial distance $r/R$ for various values of $R/H$.
The exponential cutoff of the power-law result \eqref{infalling_nostress} is evident. 
Even for the relatively large Petri dish depth $H/R=10$ the velocity is attenuated by many orders of 
magnitude relative to the unscreened case for $r/R\sim 8$, long before the sign 
oscillations are visible.  Thus, while the corresponding evolution equation
for the infalling of two colonies inherits the sign oscillations
of the flow field \eqref{eq:complexasymptoticresult}, they appear only in the
limit of very strong vertical confinement.  The screening 
would, however, lead to very marked slowing down of the infalling
trajectories relative to the infinite-depth case, and additionally
reduce the significance of further-neighbor flows on a given swimmer
in dense surface aggregates.

\section{Conclusion}
In this paper we have comprehensively explored the flows induced when Stokes singularities are placed in a Petri dish configuration, namely in a fluid layer with a bottom no-slip boundary and a top free surface boundary. In particular, we have derived both exact integral expression and far-field approximations for the flow generated by the six primary Stokes singularities: the Stokeslet, the rotlet, the source, stresslet, rotlet dipole and source dipole. Since all Stokes singularities can be expressed as derivatives of these six singularities, we can thus can gain insight about more general flows generated in a Petri dish by particles whose free space swimming fluid velocity can be represented as a sum of Stokes singularities. In particular, since the leading order contribution to the fluid velocity for these flows is separable in $z$, the full three dimensional Stokes equations can be vertically averaged to yield a much simpler two dimensional Brinkman equation much more amenable to analytic progress. 
A good example of this technique in action is \cite{Fortune21}, where the authors modeled a circular mill as a rotlet dipole, generating a radially exponentially decaying flow with $z$ dependence $\sin{(\pi z / 2 )}$, and then solve the resulting Brinkman equation in the limit that the circular mill is away from the centre of the Petri dish by transforming to bipolar coordinates.
We expect similar simplifications to hold in the many
contexts in which experiments are carried out in
the geometry of a Petri dish.

\begin{acknowledgments}
This work was supported in part by the Engineering and Physical Sciences Research 
Council, through a Doctoral Training Fellowship (GTF), by 
EPSRC grant EP/W024012/1 (EL,GTF), the European Research Council u
nder the European Union’s Horizon 2020 Research and Innovation Programme 
(Grant No. 682754, EL),
EPSRC Established Career Fellowship EP/M017982/1, Grant No. 7523 from 
the Marine Microbiology Initiative 
of the Gordon and Betty Moore 
Foundation, and the John Templeton Foundation (REG)\end{acknowledgments}

\appendix
\begin{widetext}
\section{Appendices Integral Notation and Higher Order Repeated Reflections Solutions} \label{appendixnotation}
For clarity in Appendices \ref{rotletappendix}-\ref{sourcedipoleappendix} below, we define the functions $F_{m, \, n}$ and $G_{m, \, n}$ 
\begin{subequations}
    \begin{align}
        F_{m, \, n} &= \int^{\infty}_{0} d\lambda \, \frac{ \lambda^m J_n(\lambda \rho)}{\cosh{\lambda}} \left\{
    \begin{array}{ll}
     \sinh h\lambda \, \cosh(1-z)\lambda, & z > h, \\
     \sinh z\lambda \, \cosh(1-h)\lambda, & z < h,
     \end{array} \right. \\
     G_{m, \, n} &= \int^{\infty}_{0} d\lambda \, \frac{\lambda^m J_n(\lambda \rho)}{\cosh{\lambda }} \left\{
    \begin{array}{ll}
     \sinh h\lambda \, \sinh(1-z)\lambda, & z > h, \\
     -\cosh z\lambda \, \cosh(1-h)\lambda, & z < h.
     \end{array} \right.
    \end{align}
\end{subequations}
Shown in more detail elsewhere \cite{Fortune22}, these functions allow \eqref{eq:repeatedreflection1r} in the main text to be extended to obtain repeated reflection solutions at third and fifth order,
\begin{subequations}
\begin{align}
    \mathcal{L}\left( \frac{1}{r^3} \right) = - \frac{1}{\rho}\frac{\partial}{\partial \rho}\mathcal{L}\left( \frac{1}{r} \right) &= \frac{2 F_{1, \, 1}}{\rho}, \quad \mathcal{L}\left( \frac{z}{r^3} \right) = 2 G_{1, \, 0}, \quad \mathcal{L}\left( \frac{z^2}{r^3} \right) = 2 (F_{0, \, 0} - \rho F_{1, \, 1}).\\
\mathcal{L}\left( \frac{1}{r^5} \right) &= \frac{2 (2 F_{1, \, 1} - \rho F_{2, \, 0})}{3\rho^3}, \quad \mathcal{L}\left( \frac{z}{r^5} \right) = \frac{2 G_{2, \, 1}}{3\rho}, \nonumber\\ 
\mathcal{L}\left( \frac{z^2}{r^5} \right) &= \frac{2 (\rho F_{2, \, 0} + F_{1, \, 1})}{3\rho}, \quad \mathcal{L}\left( \frac{z^3}{r^5} \right) = \frac{2 (3 G_{1, \, 0} - \rho G_{2, \, 1})}{3}.
\end{align}
\end{subequations}

\section{Rotlet in a Petri Dish} \label{rotletappendix}
The approach for a rotlet ($\epsilon_{jkp}x_p/r^3$) 
follows the procedure for the Stokeslet, with a repeated reflection solution 
\begin{equation}
    v^k_j = \mathcal{L}\left( \epsilon_{jkp}\frac{x_p}{r^3} \right) = x_{\alpha}\epsilon_{jk\alpha}\mathcal{L}\left(\frac{1}{r^3}\right) + \epsilon_{jk3}\mathcal{L}\left( \frac{z}{r^3} \right) = \frac{2 \epsilon_{jk\alpha} x_{\alpha}F_{1, \, 1}}{\rho} + 2 \epsilon_{jk3}G_{1, \, 0},
\end{equation}
with the summation convention for $\alpha \in [1 \, , \, 2]$. 
The boundary conditions for the auxiliary solution $w^k_j$ and transformed auxiliary solution $\hat{w}^k_j$ become 
\begin{subequations}
\begin{align}
    w_j^k \Big{|}_{z = 0} &= 2 \epsilon_{jk3} \int^{\infty}_{0} \lambda d\lambda \, J_0(\lambda\rho) \frac{\cosh{(1 - h)\lambda}}{\cosh{\lambda }} \Longrightarrow \hat{w}_j^k \Big{|}_{z = 0}=4\pi\epsilon_{jk3} \frac{\cosh{k(1 - h)}}{\cosh{k}}, \\
    \frac{\partial w^k_{\alpha}}{\partial z} \Big{|}_{z = 1} &= 2\epsilon_{\alpha k3} \int^{\infty}_{0} \lambda d\lambda \, J_0(\lambda\rho) \frac{\lambda \sinh{h\lambda}}{\cosh{\lambda }}
    \Longrightarrow \frac{\partial \hat{w}^k_{\alpha}}{\partial z} \Big{|}_{z = 1}=4\pi\epsilon_{\alpha k3} \frac{k \sinh{hk}}{\cosh{k}}, \\
    w^k_3 \Big{|}_{z = 1} &= - \frac{2\epsilon_{3k\alpha}x_{\alpha}}{\rho} \int^{\infty}_{0} \lambda d\lambda \, J_1(\lambda\rho) \frac{\sinh{h\lambda}}{\cosh{\lambda }}\Longrightarrow \hat{w}^k_3 \Big{|}_{z = 1}= 4\pi i k_{\alpha}\epsilon_{3 k \alpha} \frac{\sinh{hk}}{k \cosh{k}},
\end{align}
\end{subequations}
When k = 3 the boundary conditions are zero and $\hat{w}^3_j = w^3_j = 0$. When $k = \alpha \in [1 \, , \, 2]$, we find
\begin{align}
\hat{w}^{\alpha}_3 &= \frac{4\pi k_{\gamma}i \epsilon_{\gamma \alpha 3}}{k\cosh{k}(\sinh{2k} - 2k)}\Big{(} 2k\cosh{k(1-h)}\sinh{k z} - k z \cosh{k z}\sinh{k(1-h)} \nonumber \\
&- k z \sinh{k(1-h)}\cosh{k(2-z)} - 2\cosh{k}\sinh{hk}\sinh{k z}\Big{)}, \\
\hat{w}^{\alpha}_{\alpha} &= \frac{4\pi k_{\beta}k_{\alpha}\epsilon_{\beta \alpha 3} }{k\cosh{k}(\sinh{2k} - 2k)}\Big{(} z \sinh{k(h-1)}\sinh{k(2-z)} + (2 - z)\sinh(k z)\sinh{k(h-1)}\Big{)}, \\
\hat{w}^{\alpha}_{\beta} &= \frac{4\pi k_{\beta}^2 \, \epsilon_{\beta \alpha 3} }{k\cosh{k}(\sinh{2k} - 2k)}\Big{(} z \sinh{k(h-1)}\sinh{k(2-z)} + (2 - z)\sinh(k z)\sinh{k(h-1)}\Big{)} \nonumber \\
&+ \frac{4\pi\epsilon_{\beta \alpha 3} }{\cosh{k}(\sinh{2k} - 2k)}\Big{(} \sinh{2k} \cosh{k(h+z-1)} - 2k \cosh{k(h+z-1)} \Big{)},
\end{align}
where $\beta \in [1 \, , \, 2]$ and $\beta \neq \alpha$. Rewriting inverse Fourier transforms in terms of Hankel transforms, we find
\begin{subequations}
\begin{align}
    w_3^{\alpha} &= \frac{i x_{\alpha}}{2\pi \rho}\mathcal{H}_1\left( \frac{k}{k_{\alpha}}\hat{w}^{\alpha}_{3} \right), \quad w_{\alpha}^{\alpha} = -\frac{x_{\alpha} x_{\beta}}{\pi\rho^3} \mathcal{H}_1\left( \frac{k}{k_{\alpha} k_{\beta} }\hat{w}^{\alpha}_{\alpha} \right) + \frac{x_{\alpha}x_{\beta}}{2\pi \rho^2}  \mathcal{H}_0\left( \frac{k^2}{k_{\alpha} k_{\beta} }\hat{w}^{\alpha}_{\alpha} \right), \\
    w_{\beta}^{\alpha} &= \frac{1}{2\pi\rho}\left( 1 - \frac{2x_{\beta}^2}{\rho^2} \right) \mathcal{H}_1\left( k\hat{w}_1 \right) + \frac{x_{\beta}^2}{2\pi \rho^2}  \mathcal{H}_0\left(k^2\hat{w}_1 \right) + \frac{1}{2\pi}\mathcal{H}_0\left( \hat{w}_0 \right),
\end{align}
\end{subequations}
where $\beta \in [1 \, , \, 2]$, $\beta \neq \alpha$ and for notational simplicity we have decomposed $w^{\alpha}_{\beta}$ as $\hat{w}^{\alpha}_{\beta} = \hat{w}_0 + k^2_{\beta} \hat{w}_1$. 
Using contour integration, as with the Stokeslet we find
$F$ has poles of order 1 at both $z = \pi i(n + 1/2)$ ($n \in \mathbb{Z}^{\geq}$) and $z = z_0/2$ where $z_0$ satisfies $\sinh{z_0} = z_0$. When $k = 3$, $\hat{w}^3_j = w^3_j = 0$, and thus the flow field $u_j^3$ satisfies
\begin{align}
    u^3_j = v^3_j &= \frac{2\pi x_{\alpha}\epsilon_{j 3 \alpha}}{\rho} \sum_{n = 1, 3, 5, \ldots}^{\infty} n\sin{\left( \frac{n \pi h}{2} \right)}\sin{\left( \frac{n \pi z}{2} \right)} K_1\left( \frac{n \pi\rho}{2} \right).
\end{align}
Hence in the far-field, the leading order contribution decays exponentially as
\begin{equation}
    u^3_{j} = \mathcal{O}\left( \frac{\epsilon_{j3\alpha} x_{\alpha} e^{-\rho\pi/2}}{\rho^{3/2}} \right).
\end{equation}
Since $\int_{\gamma_{\epsilon}}$ vanishes as $\epsilon \rightarrow 0$, when $j = 3$ and $k = \alpha$ where $\alpha \in [1 \, , \, 2]$  (\ref{eq:residuetheorem}) simplifies to 
\begin{align}
w^{\alpha}_3 =& \frac{2\pi x_{\gamma}\epsilon_{\gamma\alpha 3}}{\rho} \sum_{n = 1, 3, 5, \ldots}^{\infty} n \sin{\left( \frac{n \pi h}{2} \right)} \sin{\left( \frac{n \pi z}{2} \right)} K_1\left( \frac{n \pi\rho}{2} \right) \nonumber \\
&- \sum_{z_0 \in \mathbb{H} \colon z_0 = \sinh{z_0}} \frac{x_{\alpha}z_0 \, H_1^1\left(\frac{\rho z_0}{2}\right)}{8\pi\rho (\cosh{z_0}-1)} \left( (\sinh{2k} - 2k)\frac{\hat{w}^{\alpha}_3}{k_{\alpha}} \right)\Bigg{|}_{k = z_0/2}, \\
u_3^{\alpha} =& v_3^{\alpha} + w_3^{\alpha} = - \sum_{z_0 \in \mathbb{H} \colon z_0 = \sinh{z_0}} \frac{x_{\alpha}z_0 \, H_1^1\left(\frac{\rho z_0}{2}\right)}{8\pi\rho (\cosh{z_0}-1)} \left( (\sinh{2k} - 2k)\frac{\hat{w}^{\alpha}_3}{k_{\alpha}} \right)\Bigg{|}_{k = z_0/2}, 
\end{align}
The contribution from poles of order 1 at $z = \pi i(n + 1/2), \, n \in \mathbb{Z}^{\geq}$ cancels out with $v_3^{\alpha}$, yielding
\begin{equation}
    u^{\alpha}_{3} = \mathcal{O}\left( \frac{\epsilon_{\gamma\alpha 3} x_{\gamma} e^{-\rho y_1/2}}{\rho^{3/2}} \right).
\end{equation}
Finally, when $j, k \in [1 \, , \, 2]$, the leading order contribution in the far-field arises from $\gamma_{\epsilon}$ i.e.
\begin{equation}
    w_{\alpha}^{\alpha} = z(2-z)\left[\epsilon_{\beta\alpha 3}\frac{6 x_{\alpha} x_{\beta} (1-h)}{\rho^4}\right], \ \ \ \ \ 
    w_{\beta}^{\alpha} = z(2-z)\left[ -\epsilon_{\beta\alpha 3} \frac{3 (1-h)}{\rho^2}\left( 1 - \frac{2 x_{\beta}^2}{\rho^2} \right) \right],
\end{equation}
where $\beta \in [1 \, , \, 2]$ and $\beta \neq \alpha$.

\section{Stresslet in a Petri Dish} \label{stressletappendix}
While the most general stresslet form is $\{x_jx_kx_l/r^5\}$, for
swimming microorganisms typically $k = l$. From fifth order repeated reflection solutions, $v_j^{k, \, l}$ for a stresslet is
\begin{align}
&v^{k, \, l}_j = \delta_{j \alpha}\delta_{k \beta}\delta_{l \delta}x_{\alpha}x_{\beta}x_{\delta}\mathcal{L}\left(\frac{1}{r^5}\right) + \delta_{j3}\delta_{k3}\delta_{l3}\mathcal{L}\left( \frac{z^3}{r^5} \right) + \left( \delta_{j \alpha}\delta_{k\beta}\delta_{l3} + \delta_{j\alpha}\delta_{k3}\delta_{l\beta} + \delta_{j3}\delta_{k\alpha}\delta_{l\beta} \right)x_{\alpha}x_{\beta}\mathcal{L}\left( \frac{z}{r^5} \right) \nonumber \\
&+ \left( \delta_{j\alpha}\delta_{k3}\delta_{l3} + \delta_{j3}\delta_{k\alpha}\delta_{l3} + \delta_{j3}\delta_{k3}\delta_{l\alpha} \right)x_{\alpha}\mathcal{L}\left( \frac{z^2}{r^5} \right) \nonumber \\
&= \frac{2x_{\alpha} F_{1,\, 1}}{3\rho}\Big{(} \delta_{j\alpha}\delta_{k3}\delta_{l3} + \delta_{j3}\delta_{k\alpha}\delta_{l3} + \delta_{j3}\delta_{k3}\delta_{l\alpha} + \frac{2x_{\beta}x_{\delta}}{\rho^2}\delta_{j\alpha}\delta_{k\beta}\delta_{l\delta} \Big{)}  \nonumber \\
&+ \frac{2x_{\alpha} F_{2, \, 0}}{3}\Big{(} \delta_{j\alpha}\delta_{k3}\delta_{l3} + \delta_{j3}\delta_{k\alpha}\delta_{l3} + \delta_{j3}\delta_{k3}\delta_{l\alpha} - \frac{x_{\beta}x_{\delta}}{\rho^2}\delta_{j\alpha}\delta_{k\beta}\delta_{l\delta} \Big{)} \nonumber \\
&+ \frac{2\rho G_{2, \, 1}}{3}\Big{(} \frac{x_{\alpha}x_{\beta}}{\rho^2}\left(\delta_{j\alpha}\delta_{k\beta}\delta_{l3} + \delta_{j\alpha}\delta_{k3}\delta_{l\beta} + \delta_{j3}\delta_{k\alpha}\delta_{l\beta} \right) - \delta_{j3}\delta_{k3}\delta_{l3} \Big{)} + 2\delta_{j3}\delta_{k3}\delta_{l3} G_{1, \, 0}, 
\end{align}
where $\{\alpha, \beta, \delta \} \in [1 \, , \, 2]$. The boundary conditions for the transformed auxiliary solution $\hat{w}^{k, \, l}_j$ 
simplify to
\begin{subequations}
\begin{align}
    \hat{w}_j^{k,\, l} \Big{|}_{z = 0} &= \frac{4\pi}{3}\left( \delta_{j\alpha}\delta_{k\beta}\delta_{l3} + \delta_{j\alpha}\delta_{k3}\delta_{l\beta} + \delta_{j3}\delta_{k\alpha}\delta_{l\beta} \right)\Bigg{(}\delta_{\alpha\beta}\frac{\cosh{k(1-h)}}{\cosh{k}} + \frac{k_{\alpha}k_{\beta}}{k}\frac{\partial}{\partial k}\left( \frac{\cosh{k(1-h)}}{\cosh{k}} \right)\Bigg{)} \nonumber \\
    &+ \frac{4\pi}{3}\delta_{j3}\delta_{k3}\delta_{l3}\Bigg{(} \frac{\cosh{k(1-h)}}{\cosh{k}} - k \frac{\partial}{\partial k}\left( \frac{\cosh{k(1-h)}}{\cosh{k}} \right)\Bigg{)}, \label{eq:stresslettransformedbc1}\\
    \frac{\partial \hat{w}^{k, \, l}_{\alpha}}{\partial z} \Bigg{|}_{z = 1} &= \frac{4\pi}{3}\left( \delta_{k\beta}\delta_{l3} + \delta_{k3}\delta_{l\beta} \right) \Bigg{(} k\delta_{\alpha\beta}\frac{\sinh{hk}}{\cosh{k}} + \frac{k_{\alpha}k_{\beta}}{k}\frac{\partial}{\partial k}\left( k \frac{\sinh{hk}}{\cosh{k}} \right) \Bigg{)}, \label{eq:stresslettransformedbc2} \\
    \hat{w}^{k, \, l}_3 \Big{|}_{z = 1} &= -\frac{4\pi i k_{\alpha}}{3}\left(\delta_{k\alpha}\delta_{l3} + \delta_{k3}\delta_{l\alpha}\right) \frac{\partial}{\partial k}\left(\frac{\sinh{hk}}{\cosh{k}}\right). \label{eq:stresslettransformedbc3}
\end{align}
\end{subequations}
As in the main text for a source, we can thus solve for $\hat{w}^{k, \, l}_j$ to give
\begin{subequations}
\begin{align}
\hat{w}^{3, \, 3}_3 &= \frac{4\pi}{3\left(\sinh{2k} - 2k\right)}\nonumber \left( \frac{\cosh{k(1-h)}}{\cosh{k}} - k\frac{\partial}{\partial k}\left( \frac{\cosh{k(1-h)}}{\cosh{k}} \right)\right) \\
&\times \Big{(} k(z - 2)\cosh{k z} + k z \cosh{k(2-z)} + \sinh{k(2-z)} - \sinh{k z} \Big{)}, \nonumber \\
\hat{w}^{3, \, 3}_{\alpha} &= \frac{4\pi k_{\alpha} i}{3\left(\sinh{2k} - 2k\right)}\left( \frac{\cosh{k(1-h)}}{\cosh{k}} - k\frac{\partial}{\partial k}\left( \frac{\cosh{k(1-h)}}{\cosh{k}} \right)\right) \nonumber \\
&\times \Big{(} (z - 2)\sinh{k z} - z\sinh{k(2-z)} \Big{)},\\
\hat{w}^{\beta, \, \delta}_3 &= \frac{4\pi}{3\left(\sinh{2k} - 2k\right)}\Bigg{(} \delta_{\beta\delta}\frac{\cosh{k(1-h)}}{\cosh{k}} + \frac{k_{\beta}k_{\delta}}{k}\frac{\partial}{\partial k}\left( \frac{\cosh{k(1-h)}}{\cosh{k}} \right)\Bigg{)} \nonumber \\
&\times \Big{(} k(z - 2)\cosh{k z} + k z \cosh{k(2-z)} + \sinh{k(2-z)} - \sinh{k z} \Big{)}, \\
\hat{w}^{\beta, \, \delta}_{\alpha} &= \frac{4\pi k_{\alpha} i}{3\left(\sinh{2k} - 2k\right)}\Bigg{(} \delta_{\beta\delta}\frac{\cosh{k(1-h)}}{\cosh{k}} + \frac{k_{\beta}k_{\delta}}{k}\frac{\partial}{\partial k}\left( \frac{\cosh{k(1-h)}}{\cosh{k}} \right) \Bigg{)} \nonumber \\
&\times \Big{(} (z - 2)\sinh{k z} - z\sinh{k(2-z)} \Big{)}, \\
\hat{w}^{\alpha, \, 3}_3 &= \frac{8\pi ik_{\alpha}}{3\cosh^2{k}\left( \sinh{2k} - 2k \right)}\Big{(} hk z\cosh^3{k}\cosh{k\left(1-h-z\right)} - hk\cosh{k}\sinh{k z}\sinh{k(1-h)} \nonumber \\
&- k\sinh{hk}\sinh{k z} - k z\cosh{k}\sinh{hk}\sinh{k(1-z)} - hk z\cosh^2{k}\sinh{hk}\sinh{k z} \nonumber \\
&- z\cosh^2{k}\sinh{k}\cosh{k(1-h-z)} + \cosh{k}\sinh{k z}\cosh{k(1+h)} \nonumber \\
&- h\cosh^2{k}\cosh{h k}\sinh{k z}\Big{)}, \\
\hat{w}^{\alpha, \, 3}_{\beta} &= \frac{8\pi k_{\alpha}k_{\beta}}{3k\cosh^2{k}\left( \sinh{2k} - 2k \right)}\Big{(} k\sinh{hk}\cosh{k z} - k z \cosh{k}\sinh{hk}\cosh{k(1-z)} \nonumber \\
&+ hk\cosh{k}\cosh{k z}\sinh{k(1-h)} + hk z \cosh^2{k}\sinh{hk}\cosh{k z} + hk z \cosh^3{k}\sinh{k(1-h-z)} \nonumber \\
&- \cosh{k}\sinh{k}\sinh{k(h+z)} - (h+z)\cosh^2{k}\sinh{k}\sinh{k(1-h-z)} \Big{)} \nonumber \\
&+ \frac{4\pi \delta_{\alpha\beta}}{3} \cosh{k}\cosh{k(1-h-z)}, \label{eq:stresslet132}
\end{align}
\end{subequations}
where $\alpha, \, \beta, \delta \in [1, \, 2]$. Hence, as above, we find the following integral expressions for $w_j^{k, \, l}$.
\begin{subequations}
\begin{align}
    w_3^{3, \, 3} &= \frac{1}{2\pi}\mathcal{H}_0\left(\hat{w}^{3, \, 3}_3\right), \quad w_{\alpha}^{3, \, 3} = \frac{i x_{\alpha}}{2\pi \rho}\mathcal{H}_1\left( \frac{k}{k_{\alpha}}\hat{w}^{3, \, 3}_{\alpha} \right), \quad w_{3}^{\alpha, \, 3} = \frac{i x_{\alpha}}{2\pi \rho}\mathcal{H}_1\left( \frac{k}{k_{\alpha}}\hat{w}^{\alpha, \, 3}_{3} \right), \\
    w_{3}^{\beta, \, \delta} &= \frac{\delta_{\beta\delta}}{2\pi}\mathcal{H}_0\left( \hat{w}^{\beta, \, \delta}_{3, \, 1} \right) + \frac{x_{\beta}x_{\delta}}{2\pi\rho^2}\mathcal{H}_0 \left( k^2 \hat{w}^{\beta, \, \delta}_{3, \, 2} \right) + \frac{1}{2\pi\rho}\left( \delta_{\beta\delta} - \frac{2x_{\beta}x_{\delta}}{\rho^2} \right)\mathcal{H}_1\left( k \, \hat{w}^{\beta, \, \delta}_{3, \, 2} \right), \label{eq:stresshankelw3betadelta} \\
    w_{\alpha}^{\beta, \, \delta} &= \frac{i \delta_{\beta\delta} x_{\alpha}}{2\pi \rho}\mathcal{H}_1\left(k\hat{w}^{\beta, \, \delta}_{\alpha, \, 1} \right) + \frac{i \, x_{\alpha}x_{\beta}x_{\delta}}{2\pi\rho^3} \, \mathcal{H}_1\left( k^3 \hat{w}^{\beta, \, \delta}_{\alpha, \, 2} \right) \nonumber \\
    &+ \frac{i}{2\pi\rho^3}\left( x_{\alpha}\delta_{\beta\delta} + x_{\beta}\delta_{\alpha\delta} + x_{\delta}\delta_{\alpha\beta} - \frac{4x_{\alpha}x_{\beta}x_{\delta}}{\rho^2} \right)\left( 2 \mathcal{H}_1\left(k \, \hat{w}_{\alpha, \, 2}^{\beta, \, \delta}\right) - \rho \mathcal{H}_0\left( k^2\hat{w}^{\beta, \, \delta}_{\alpha, \, 2} \right)  \right), \\
    w_{\beta}^{\alpha, \, 3} &= \frac{1}{2\pi}\mathcal{H}_0\left( \hat{w}^{\alpha, \, 3}_{\beta, \, 1} \right) + \frac{x_{\alpha}x_{\beta}}{2\pi\rho^2}\mathcal{H}_0 \left( k^2 \hat{w}^{\alpha, \, 3}_{\beta, \, 2} \right) + \frac{1}{2\pi\rho}\left( \delta_{\alpha\beta} - \frac{2x_{\alpha}x_{\beta}}{\rho^2} \right)\mathcal{H}_1\left( k \, \hat{w}^{\alpha, \, 3}_{\beta, \, 2} \right),
\end{align}
\end{subequations}
where for notational simplicity, we have decomposed $\hat{w}^{\beta, \, \delta}_{3}, \hat{w}^{\beta, \, \delta}_{\alpha}$ and $\hat{w}^{\alpha, \, 3}_{\beta}$ as 
\begin{equation}
\hat{w}^{\beta, \, \delta}_{3} = \hat{w}^{\beta, \, \delta}_{3, \, 1} + k_{\beta}k_{\delta}\hat{w}^{\beta, \, \delta}_{3, \, 2}, \quad \hat{w}^{\beta, \, \delta}_{\alpha} = k_{\alpha}\hat{w}^{\beta, \, \delta}_{\alpha, \, 1} + k_{\alpha}k_{\beta}k_{\delta}\hat{w}^{\beta, \, \delta}_{\alpha, \, 2}, \quad \hat{w}^{\alpha, \, 3}_{\beta} = \hat{w}^{\alpha, \, 3}_{\beta, \, 1} + k_{\alpha}k_{\beta}\hat{w}^{\alpha, \, 3}_{\beta, \, 2}.
\end{equation}
Similarly to the source above, F has in $\gamma$ poles of order 2 at $z = \pi i(n + 1/2)$ where $n \in \mathbb{Z}^{\geq}$ and poles of order 1 at $z = z_0/2$ where $z_0$ satisfies $\sinh{z_0} = z_0$. Since $\int_{\gamma_{\epsilon}}$ vanishes as $\epsilon \rightarrow 0$, when $j = k = l = 3$ (\ref{eq:residuetheorem}) simplifies to become
\begin{align}
w^{3, \, 3}_3 =& \frac{2\pi}{3} \sum_{n = 1, 3, 5, \ldots}^{\infty} n\cos{\left(\frac{n \pi z}{2}\right)}\sin{\left(\frac{n \pi h}{2}\right)} \Bigg{(}3 K_0\left(\frac{n \pi\rho}{2}\right) - \frac{n \pi\rho}{2}K_1\left( \frac{n \pi\rho}{2} \right) \Bigg{)} \nonumber \\
&+ \sum_{z_0 \in \mathbb{H} \colon z_0 = \sinh{z_0}} \frac{i z_0}{8(\cosh{z_0}-1)} \Big{(} \hat{w}^{3, \, 3}_3 (\sinh{2k} - 2k) \Big{)}\Big{|}_{k = z_0/2} H_0^1\left(\frac{\rho z_0}{2}\right), \\
u_3^{3, \, 3} =& v_3^{3, \, 3} + w^{3, \, 3}_3 = \mathcal{O}\left( \frac{e^{-\rho y_1/2}}{\sqrt{\rho}}  \right),
\end{align}
noting that as for the Stokeslet, the contribution from the poles of order $2$ in $w^{3,\, 3}_3$ cancels out with $v^{3, \, 3}_3$. Similarly, the leading order contribution in the far-field when $j = 3$ for the other cases for $k$ and $l$ are
\begin{equation}
u^{3, \, \alpha}_3 = u^{\alpha, \, 3}_3  = \mathcal{O}\left( \frac{x_{\alpha} \, e^{-\rho y_1/2}}{\rho^{3/2}} \right), \,
    u^{\beta, \, \delta}_3 = \mathcal{O}\left( \delta_{\beta\delta}\frac{e^{-\rho y_1/2}}{\sqrt{\rho}} \right) + \mathcal{O}\left( x_{\beta}x_{\delta}\frac{e^{-\rho y_1/2}}{\rho^{5/2}} \right).
\end{equation}
Finally, when $j = \alpha \in [1 \, , \, 2]$, the leading order contribution in the far-field arises from $\gamma_{\epsilon}$, namely
\begin{subequations}
\begin{align}
    u^{3, \, 3}_{\alpha} &= \frac{z x_{\alpha}}{\rho^2}\left(2 - z \right), \quad u^{\alpha, \, 3}_{\beta} = z(2-z)\left[  - \frac{1-h}{\rho^2}\left( \delta_{\alpha\beta} - \frac{2x_{\alpha}x_{\beta}}{\rho^2} \right) \right]. \\
    u^{\beta, \, \delta}_{\alpha} &= z\left(2 - z\right)\Bigg{(}\frac{x_{\alpha}}{\rho^2}\delta_{\beta\delta} - \frac{2h}{\rho^4}\left( 2 - h \right) \left(x_{\alpha}\delta_{\beta\delta} + x_{\beta}\delta_{\alpha\delta} + x_{\delta}\delta_{\alpha\beta} - \frac{4x_{\alpha}x_{\beta}x_{\delta}}{\rho^2}\right) \Bigg{)}, 
\end{align}
\end{subequations}

\section{Rotlet Dipole in a Petri Dish} \label{rotletdipoleappendix}
From the fifth order repeated reflection solutions (Appendix \ref{appendixnotation}), $v_j^{k}$ for a rotlet dipole is
\begin{align}
    v^k_j &= \mathcal{L}\left( \epsilon_{jpk}\frac{x_p x_k}{r^5} \right) = \delta_{j3}\delta_{k\alpha}x_{\alpha}x_{\beta}\epsilon_{3\beta\alpha}\mathcal{L}\left(\frac{1}{r^5}\right) + \delta_{j \alpha}x_{\beta}\left( \delta_{k3}\epsilon_{\alpha \beta 3} + \delta_{k \beta}\epsilon_{\alpha 3 \beta} \right)\mathcal{L}\left( \frac{z}{r^3} \right) \nonumber \\
     &= \epsilon_{3\beta\alpha} \, \frac{4\delta_{j3}\delta_{k\alpha}x_{\alpha}x_{\beta} F_{1, \, 1}}{3\rho^3} - \epsilon_{3\beta\alpha} \, \frac{2\delta_{j3}\delta_{k\alpha}x_{\alpha}x_{\beta} F_{2, \, 0}}{3\rho^2} + \frac{2 \delta_{j \alpha} x_{\beta}G_{2, \, 1}}{3\rho} \left( \delta_{k3}\epsilon_{\alpha \beta 3} + \delta_{k \beta}\epsilon_{\alpha 3  \beta} \right),
\end{align}
with boundary conditions for the corresponding auxiliary solution $w^k_j$ and transformed auxiliary solution $\hat{w}^k_j$
\begin{subequations}
\begin{align}
    w_j^k \Big{|}_{z = 0} &= \frac{2\delta_{j \alpha}x_{\beta}}{3\rho} \left( \delta_{k3}\epsilon_{\alpha \beta 3} + \delta_{k \beta} \epsilon_{\alpha 3 \beta} \right) \int^{\infty}_{0} d\lambda \, J_1(\lambda\rho) \frac{\lambda^2 \cosh{(1 - h)\lambda}}{\cosh{\lambda }} \Longrightarrow\nonumber \\
    &\hat{w}_j^k \Big{|}_{z = 0}= -\frac{4 \pi i \delta_{j \alpha} k_{\beta}}{3} \left( \delta_{k3}\epsilon_{\alpha \beta 3} + \delta_{k \beta}\epsilon_{\alpha 3 \beta} \right) \frac{\cosh{k(1 - h)}}{\cosh{k}}, \label{eq:rotletdipolebc1} \\
    \frac{\partial w^k_{\alpha}}{\partial z} \Big{|}_{z = 1} &= \frac{2x_{\beta}}{3\rho} \left(\delta_{k3}\epsilon_{\alpha\beta 3} + \delta_{k \beta}\epsilon_{\alpha 3 \beta} \right) \int^{\infty}_{0} d\lambda \, J_1(\lambda\rho) \frac{\lambda^3 \sinh{h\lambda}}{\cosh{\lambda }} \Longrightarrow \nonumber \\
    \frac{\partial \hat{w}^k_{\alpha}}{\partial z} \Big{|}_{z = 1}&= -\frac{4\pi k_{\beta}}{3} \left( \delta_{k3}\epsilon_{\alpha \beta 3} + \delta_{k \beta}\epsilon_{\alpha 3 \beta} \right) \frac{k \sinh{hk}}{\cosh{k}}, \label{eq:rotletdipolebc2}\\
    w^k_3 \Big{|}_{z = 1} &= - \frac{4\delta_{k\alpha}x_{\alpha}x_{\beta}}{3\rho^3}\epsilon_{3 \beta \alpha} \int^{\infty}_{0} d\lambda \, J_1(\lambda\rho) \frac{\lambda \sinh{h\lambda}}{\cosh{\lambda }} + \frac{2\delta_{k\alpha}x_{\alpha}x_{\beta}}{3\rho^2}\epsilon_{3\beta\alpha}\int^{\infty}_{0} d\lambda \, J_0(\lambda\rho) \frac{\lambda^2 \sinh{h\lambda}}{\cosh{\lambda }}. \label{eq:rotletdipolebc3}
\end{align}
\end{subequations}
However, (\ref{eq:rotletdipolebc3}) is difficult to transform. Noting that $\alpha \neq \beta$ and utilising Bessel function identities, we find
\begin{align}
    \hat{w}_3^k \Big{|}_{z = 1} &= -\frac{2 \delta_{k \alpha} \epsilon_{3 \beta \alpha}}{3}\frac{k_{\alpha}k_{\beta}}{k}\frac{\partial}{\partial k}\left( \frac{1}{k} \left( 2\pi \int^{\infty}_0 \int^{\infty}_0 d\rho d\lambda \frac{\sinh{h \lambda}}{\cosh{\lambda }} J_0{k \rho} \left( \frac{2 \lambda J_1{\lambda \rho}}{\rho^2} - \frac{\lambda^2 J_0(\lambda \rho)}{\rho}\right)\right)\right) \nonumber \\
    &= -\frac{4\pi \delta_{k \alpha} \epsilon_{3 \beta \alpha}}{3}\frac{k_{\alpha}k_{\beta}}{k}\frac{\partial}{\partial k}\left( \frac{1}{k} \left( g_1 - g_2 \right) \right), \label{eq:messyform}
\end{align}
where $g_1$ and $g_2$ are defined as satisfying respectively
\begin{equation}
    g_1 = \int^{\infty}_{0} \int^{\infty}_{0} d\rho d\lambda \frac{\lambda^2 \sinh{h \lambda}}{\cosh{\lambda }} J_0(\lambda \rho) J_1{(k \rho)}, \quad g_2 = \int^{\infty}_{0}\int^{\infty}_0 d\rho d\lambda \frac{2 \lambda \sinh{(h \lambda)}}{\cosh{\lambda }} \frac{J_1(\lambda \rho) J_1(k \rho)}{\rho}.
\end{equation}
However, $g_1$ simplifies to give
\begin{equation}
    g_1 = \int^\infty_0 \frac{\lambda^2 \sinh{h\lambda}}{\cosh{\lambda }}\left[ \int^{\infty}_0 \rho d\rho J_0(\rho \lambda) \left( \frac{J_1(\rho\lambda)}{\rho} \right) \right] = \frac{1}{k}\int^{k}_0 d\lambda \frac{\lambda^2 \sinh{h\lambda}}{\cosh{\lambda }}.
\end{equation}
Furthermore, $g_2$ simplifies to give
\begin{align}
    g_2 &= \int^{\infty}_0 d\lambda \frac{2\lambda \sinh{2h\lambda}}{\cosh{\lambda }}\left[ \int^{\infty}_0 d\rho \frac{J_1(k\rho)J_2(\lambda \rho)}{\rho} \right] = \int^{\infty}_{0} d\lambda \frac{2\lambda \sinh{h\lambda}}{\cosh{\lambda }} \left[ \frac{\lambda k}{k^2 + \lambda^2 + |k^2 - \lambda^2| } \right] \nonumber \\
    &= \int^{k}_{0} d\lambda \frac{\lambda^2}{k}\frac{\sinh{h \lambda}}{\cosh{\lambda }} + \int^{\infty}_k d\lambda \frac{k \sinh{h\lambda}}{\cosh{\lambda }}.
\end{align}
Putting this all together, (\ref{eq:messyform}) becomes
\begin{equation}
\hat{w}_3^k \Big{|}_{z = 1} = \frac{4\pi \delta_{k \alpha} \epsilon_{3 \beta \alpha}}{3}\frac{k_{\alpha}k_{\beta}}{k}\left[ \frac{\sinh{hk}}{\cosh{k }} \right].
\end{equation}
Hence, as in the main text for a source, we can thus solve for $\hat{w}^k_j$ to give
\begin{subequations}
\begin{equation}
\hat{w}_3^3 = 0, \quad \hat{w}^{3}_{\alpha} = \frac{-4\pi k_{\beta} i \epsilon_{\alpha \beta 3}}{3}\frac{\cosh{k(1 - h - z)}}{\cosh{k }},
\end{equation}
\begin{align}
\hat{w}^{\alpha}_3 &= -\frac{4\pi k_{\alpha} k_{\beta} i \epsilon_{\beta \alpha 3}}{3k\cosh{k}(\sinh{2k} - 2k)}\Big{(} 2k\cosh{k(1-h)}\sinh{k z} - k z \cosh{k z}\sinh{k(1-h)} \nonumber \\
&- k z \sinh{k(1-h)}\cosh{k(2-z)} - 2\cosh{k}\sinh{hk}\sinh{k z}\Big{)}, \\
\hat{w}^{\alpha}_{\alpha} &= \frac{4\pi k_{\alpha}^2k_{\beta}\, i\epsilon_{\beta \alpha 3} }{3k\cosh{k}(\sinh{2k} - 2k)}\Big{(} z \sinh{k(h-1)}\sinh{k(2-z)} + (2 - z)\sinh(k z)\sinh{k(h-1)}\Big{)}, \\
\hat{w}^{\alpha}_{\beta} &= \frac{4\pi k_{\alpha}k_{\beta}^2 i\, \epsilon_{\beta \alpha 3} }{3k\cosh{k}(\sinh{2k} - 2k)}\Big{(} z \sinh{k(h-1)}\sinh{k(2-z)} + (2 - z)\sinh(k z)\sinh{k(h-1)}\Big{)} \nonumber \\
&+ \frac{4\pi k_{\alpha}i\epsilon_{\beta \alpha 3} }{3\cosh{k}(\sinh{2k} - 2k)}\Big{(} \sinh{2k} \cosh{k(h+z-1)} - 2k \cosh{k(h+z-1)} \Big{)},
\end{align}
\end{subequations}
where $\beta \in [1 \, , \, 2]$ and $\beta \neq \alpha$. Rewriting the inverse Fourier transform in terms of Hankel transforms, we get the following integral expressions for $w_j^{\alpha}$
\begin{subequations}
\begin{align}
    w_3^3 &= 0, \quad w_{\alpha}^3 = \frac{i x_{\beta}}{2\pi\rho} \mathcal{H}_1\left( \frac{k}{k_{\beta}} \hat{w}^3_{\alpha} \right), \quad w_{3}^{\alpha} = -\frac{x_{\alpha} x_{\beta}}{\pi\rho^3} \mathcal{H}_1\left( \frac{k}{k_{\alpha} k_{\beta} }\hat{w}^{\alpha}_{3} \right) + \frac{x_{\alpha}x_{\beta}}{2\pi \rho^2}  \mathcal{H}_0\left( \frac{k^2}{k_{\alpha} k_{\beta} }\hat{w}^{\alpha}_{3} \right), \\
    w_{\alpha}^{\alpha} &= \frac{i x_{\alpha}^2 x_{\beta}}{2\pi\rho^3} \mathcal{H}_1\left( \frac{k^3}{k_{\alpha}^2 k_{\beta}} \hat{w}_{\alpha}^{\alpha} \right) + \frac{i}{2 \pi \rho^3}\left( x_{\beta} - \frac{4 x_{\alpha}^2 x_{\beta}}{\rho^2} \right)\left( 2 \mathcal{H}_1\left( \frac{k}{k_{\alpha}^2 k_{\beta}} \hat{w}_{\alpha}^{\alpha} \right)  - \rho \mathcal{H}_0\left( \frac{k^2}{k_{\alpha}^2 k_{\beta}} \hat{w}_{\alpha}^{\alpha} \right) \right), \\
    w_{\beta}^{\alpha} &= \frac{i x_{\alpha}}{2\pi\rho}\mathcal{H}_1 \left( k \hat{w}_0 \right) + \frac{i x_{\alpha} x_{\beta}^2}{2\pi\rho^3} \mathcal{H}_1\left( k^3\hat{w}_1 \right) + \frac{i}{2 \pi \rho^3}\left( x_{\alpha} - \frac{4 x_{\alpha} x_{\beta}^2}{\rho^2} \right)\left( 2 \mathcal{H}_1\left( k\hat{w}_1 \right)  - \rho \mathcal{H}_0\left( k^2 \hat{w}_{1} \right) \right)
\end{align}
\end{subequations}
where $\beta \in [1 \, , \, 2]$, $\beta \neq \alpha$ and for notational simplicity we have decomposed $\hat{w}^{\alpha}_{\beta}$ as $\hat{w}^{\alpha}_{\beta} = k_{\alpha}\left(\hat{w}_0 + k^2_{\beta} \hat{w}_1\right)$. When $k = j = 3$, $u_3^3 = v_3^3 = w^3_3 = 0$. Furthermore, when $k = 3$ and $j = \alpha$, we have
\begin{align}
    v^3_{\alpha} &= -\frac{ x_{\beta} \epsilon_{\alpha \beta 3}}{3\rho}\sum_{n = 1, 3, 5, \ldots}^{\infty} \pi^2 n^2 K_1\left( \frac{n \pi\rho}{2} \right)\Bigg{[} \sin{\left( \frac{n \pi h}{2} \right)}\cos{\left( \frac{n \pi z}{2} \right)} \Bigg{]} \\
     w^3_{\alpha} &= \frac{ x_{\beta} \epsilon_{\alpha \beta 3}}{3\rho}\sum_{n = 1, 3, 5, \ldots}^{\infty} \pi^2 n^2 K_1\left( \frac{n \pi\rho}{2} \right)\Bigg{[} \sin{\left( \frac{n \pi h}{2} \right)}\cos{\left( \frac{n \pi z}{2} \right)} + \cos{\left( \frac{n \pi h}{2} \right)}\sin{\left( \frac{n \pi z}{2} \right)}\Bigg{]}, \\
    u_{\alpha}^3 &= v_{\alpha}^3 + w_{\alpha}^3 = \frac{x_{\beta} \epsilon_{\alpha \beta 3}}{3\rho}\sum_{n = 1, 3, 5, \ldots}^{\infty} \pi^2 n^2 K_1\left( \frac{n \pi\rho}{2} \right)\Bigg{[} \cos{\left( \frac{n \pi h}{2} \right)}\sin{\left( \frac{n \pi z}{2} \right)} \Bigg{]}.
\end{align}
Hence in the far-field, the leading order contribution decays exponentially as
\begin{equation}
    u^3_{\alpha} = \mathcal{O}\left( \frac{\epsilon_{\alpha\beta 3} x_{\beta} e^{-\rho\pi/2}}{\rho^{3/2}} \right).
\end{equation}
Since $\int_{\gamma_{\epsilon}}$ vanishes as $\epsilon \rightarrow 0$, when $j = 3$ and $k = \alpha$ where $\alpha \in [1 \, , \, 2]$  (\ref{eq:residuetheorem}) becomes
\begin{align}
w^{\alpha}_3 =& -\frac{4 x_{\alpha}x_{\beta}\epsilon_{3\beta\alpha}}{3\rho^2} \sum_{n = 1, 3, 5, \ldots}^{\infty} \sin{\left( \frac{n \pi h}{2} \right)} \sin{\left( \frac{n \pi z}{2} \right)} K_0\left( \frac{n \pi\rho}{2} \right) \nonumber \\
&- \frac{4 \pi x_{\alpha}x_{\beta}\epsilon_{3\beta\alpha}}{3\rho^3} \sum_{n = 1, 3, 5, \ldots}^{\infty} n \sin{\left( \frac{n \pi h}{2} \right)} \sin{\left( \frac{n \pi z}{2} \right)} K_1\left( \frac{n \pi\rho}{2} \right) \nonumber \\
&- \frac{x_{\alpha}x_{\beta}}{ \rho^3} \sum_{z_0 \in \mathbb{H} \colon z_0 = \sinh{z_0}} \frac{ i z_0^2 \, H_1^1\left(\frac{\rho z_0}{2}\right)}{8(\cosh{z_0}-1)} \left( (\sinh{2k} - 2k)\frac{\hat{w}^{\alpha}_3}{k_{\alpha}} \right)\Bigg{|}_{k = z_0/2}, \nonumber \\
&+ \frac{x_{\alpha}x_{\beta}}{\rho^2} \sum_{z_0 \in \mathbb{H} \colon z_0 = \sinh{z_0}} \frac{ i z_0^3 \, H_0^1\left(\frac{\rho z_0}{2}\right)}{32(\cosh{z_0}-1)} \left( (\sinh{2k} - 2k)\frac{\hat{w}^{\alpha}_3}{k_{\alpha}} \right)\Bigg{|}_{k = z_0/2}, \\
u_3^{\alpha} =& v_3^{\alpha} + w_3^{\alpha} = - \frac{x_{\alpha}x_{\beta}}{ \rho^3} \sum_{z_0 \in \mathbb{H} \colon z_0 = \sinh{z_0}} \frac{ i z_0^2 \, H_1^1\left(\frac{\rho z_0}{2}\right)}{8(\cosh{z_0}-1)} \left( (\sinh{2k} - 2k)\frac{\hat{w}^{\alpha}_3}{k_{\alpha}} \right)\Bigg{|}_{k = z_0/2}, \nonumber \\
&+ \frac{x_{\alpha}x_{\beta}}{ \rho^2} \sum_{z_0 \in \mathbb{H} \colon z_0 = \sinh{z_0}} \frac{ i z_0^3 \, H_0^1\left(\frac{\rho z_0}{2}\right)}{32(\cosh{z_0}-1)} \left( (\sinh{2k} - 2k)\frac{\hat{w}^{\alpha}_3}{k_{\alpha}} \right)\Bigg{|}_{k = z_0/2}, 
\end{align}
noting that the contribution from the poles of order 1 at $z = \pi i(n + 1/2)$ where $n \in \mathbb{Z}^{\geq}$ cancel out with $v_3^{\alpha}$. Hence,
\begin{equation}
    u^{\alpha}_{3} = \mathcal{O}\left( \frac{\epsilon_{3 \beta\alpha} x_{\alpha}x_{\beta} e^{-\rho y_1/2}}{\rho^{5/2}} \right).
\end{equation}
Finally, when $j, k \in [1 \, , \, 2]$, the leading order contribution in the far-field arises from $\gamma_{\epsilon}$ i.e.
\begin{equation}
    w_{\alpha}^{\alpha} = z(2-z)\left[ -\epsilon_{\alpha\beta 3}\frac{2 x_{\beta} (1-h)}{\rho^4}\left( 1 - \frac{4 x_{\alpha}^2}{\rho^2} \right) \right], \quad w_{\beta}^{\alpha} = z(2-z)\left[ -\epsilon_{\alpha\beta 3} \frac{2 x_{\alpha} (1-h)}{\rho^4}\left( 1 - \frac{4 x_{\beta}^2}{\rho^2} \right) \right],
\end{equation}
where $\beta \in [1 \, , \, 2]$ and $\beta \neq \alpha$.
\section{Source Dipole in a Petri Dish} \label{sourcedipoleappendix}
Using fifth order repeated reflection solutions (Appendix \ref{appendixnotation}), $v_j^{k}$ for a source dipole becomes
\begin{align}
    v_j^k &= \delta_{jk} \mathcal{L}\left( \frac{1}{r^3} \right) - 3\delta_{j\alpha}\delta_{k\beta} x_{\alpha}x_{\beta}\mathcal{L}{\left( \frac{1}{r^5} \right)} - 3(\delta_{j\alpha}\delta_{k3} + \delta_{k\alpha}\delta_{j3})x_{\alpha}\mathcal{L}\left( \frac{z}{r^5} \right) - 3\delta_{j3}\delta_{k3}\mathcal{L}\left( \frac{z^2}{r^5} \right) \nonumber \\
    &= 2F_{2, \, 0}\left( \frac{\delta_{j\alpha}\delta_{k \beta} x_{\alpha}x_{\beta}}{\rho^2} - \delta_{j3}\delta_{k3} \right) + \frac{2 F_{1, \, 1}}{\rho} \left( \delta_{jk} - \frac{2\delta_{j\alpha}\delta_{k\beta}x_{\alpha}x_{\beta}}{\rho^2} - \delta_{j3}\delta_{k3} \right) - \frac{2x_{\alpha} G_{2, \, 1}}{\rho}\left( \delta_{j3}\delta_{k\alpha} + \delta_{k3}\delta_{j\alpha} \right). 
\end{align}
The boundary conditions for the corresponding auxiliary solution $w^k_j$ and transformed auxiliary solution $\hat{w}^k_j$ become 
\begin{subequations}
\begin{align}
    w_j^k \Big{|}_{z = 0} &= -\frac{2x_{\alpha}}{\rho} \left( \delta_{j 3} \delta_{k \alpha} + \delta_{k 3} \delta_{j \alpha} \right) \int^{\infty}_{0} \lambda^2 d\lambda \, J_1(\lambda \rho)\frac{\cosh{(1 - h)\lambda}}{\cosh{\lambda }} \Longrightarrow \nonumber \\
    \hat{w}_j^k \Big{|}_{z = 0} &= 4\pi i k_{\alpha} \left( \delta_{j 3}\delta_{k \alpha} + \delta_{k 3}\delta_{j \alpha} \right) \frac{\cosh{k(1 - h)}}{\cosh{k}}, \\
    \frac{\partial w_{\alpha}^k}{\partial z}\Bigg{|}_{z = 1} &= -\frac{2x_{\alpha} \delta_{k 3}}{\rho} \int^{\infty}_{0} \lambda^2 d\lambda \, J_1(\lambda \rho) \frac{\lambda \sinh{h\lambda}}{\cosh{\lambda }} \Longrightarrow \frac{\partial \hat{w}_{\alpha}^k}{\partial z}\Bigg{|}_{z = 1}= 4\pi i k_{\alpha} \delta_{k 3}\frac{k \sinh{hk}}{\cosh{k}}, \\
    w_3^k \Big{|}_{z = 1} &= 2\delta_{k3}\int^{\infty}_{0} \lambda d\lambda \, J_0(\lambda\rho) \frac{\lambda \sinh{h\lambda}}{\cosh{\lambda }} \Longrightarrow \hat{w}_3^k \Big{|}_{z = 1}= 4\pi \delta_{k3} \frac{k \sinh{hk}}{\cosh{k}}.
\end{align}
\end{subequations}
We thus obtain
\begin{subequations}
\begin{align}
\hat{w}^3_3 &= \frac{4\pi k}{\cosh{k}\left( \sinh{2k} - 2k \right)}\Big{(} 2\cosh{k}\sinh{hk}\sinh{k z} - 2k\sinh{k z}\cosh{k(1-h)} \nonumber \\
&+ 2k z \sinh{k(1-h)}\cosh{k }\cosh{k(1 - z)} \Big{)}, \\
\hat{w}^3_{\alpha} &= \frac{4\pi i k_{\alpha}}{\cosh{k}\left( \sinh{2k} - 2k \right)}\Big{(} \sinh{2k}\cosh{k(1-h-z)} - 2k\cosh{k z}\cosh{k(1-h)} \nonumber \\
&- 2k z \cosh{k} \sinh{k (1-z)}\sinh{k(1-h)} \Big{)}, \\
\hat{w}^{\alpha}_3 &= \frac{4\pi ik_{\alpha} \cosh{k(1-h)}}{\cosh{k}(\sinh{2k} - 2k)}\Big{(}k(z-2)\cosh{k z} + k z \cosh{k(2-z)} + \sinh{k(2-z)} - \sinh{k z}\Big{)}, \\
\hat{w}^{\alpha}_{\beta} &= \frac{4\pi k_{\alpha}k_{\beta} \cosh{k(1-h)}}{\cosh{k}(\sinh{2k} - 2k)}\left( z \sinh{k(2-z)} - (z - 2)\sinh{k z} \right),
\end{align}
\end{subequations}
or, utilizing Hankel transforms,
\begin{subequations}
\begin{align}
    w_3^3 &= \frac{1}{2\pi}\mathcal{H}_0\left(\hat{w}^3_3\right), \quad w_{\alpha}^3 = \frac{i x_{\alpha}}{2\pi \rho}\mathcal{H}_1\left( \frac{k}{k_{\alpha}}\hat{w}^3_{\alpha} \right), \quad w_{3}^{\alpha} = \frac{i x_{\alpha}}{2\pi \rho}\mathcal{H}_1\left( \frac{k}{k_{\alpha}}\hat{w}^{\alpha}_{3} \right), \\
w_{\beta}^{\alpha} &= \frac{1}{2\pi}\left( \frac{\delta_{\alpha \beta}}{\rho} - 2\frac{x_{\alpha} x_{\beta}}{\rho^3} \right) \mathcal{H}_1\left( \frac{k}{k_{\alpha} k_{\beta} }\hat{w}^{\alpha}_{\beta} \right) + \frac{x_{\alpha}x_{\beta}}{2\pi \rho^2}  \mathcal{H}_0\left( \frac{k^2}{k_{\alpha} k_{\beta} }\hat{w}^{\alpha}_{\beta} \right). 
\end{align}
\end{subequations}
Similarly to the source, $F$ has poles of order 1 at $z = \pi i(n + 1/2)$ where $n \in \mathbb{Z}^{\geq}$ and poles of order 1 at $z = z_0/2$ where $\sinh{z_0} = z_0$. When $j = k = 3$, since $\int_{\gamma_{\epsilon}}$ vanishes as $\epsilon \rightarrow 0$, and (\ref{eq:residuetheorem}) becomes
\begin{align}
 w^3_3 =& -\sum_{n = 1, 3, 5, \ldots}^{\infty} \pi^2 n^2\sin{\Bigg{(}\frac{n \pi h}{2}\Bigg{)}}\sin{\Bigg{(}\frac{n \pi z}{2}\Bigg{)}} K_0\Bigg{(}\frac{n \pi\rho}{2}\Bigg{)} \nonumber \\
&+ \sum_{z_0 \in \mathbb{H} \colon z_0 = \sinh{z_0}} \frac{i z_0}{8(\cosh{z_0}-1)} \Big{(} \hat{w}^3_3 (\sinh{2k} - 2k) \Big{)}\Big{|}_{k = z_0/2} H_0^1\left(\frac{\rho z_0}{2}\right), \nonumber \\
\Longrightarrow u_3^3 =& v_3^3 + w^3_3 = \sum_{z_0 \in \mathbb{H} \colon z_0 = \sinh{z_0}} \frac{i z_0}{8(\cosh{z_0} - 1}\Big{(} \hat{w}^3_3 (\sinh{2k} - 2k) \Big{)}\Big{|}_{k = z_0/2} H_0^1\left(\frac{\rho z_0}{2}\right),
\end{align}
Hence, in the far-field the leading order contribution to $u^3_3$ is
\begin{equation}
    u^3_3 = \mathcal{O}\left( \frac{e^{-\rho y_1/2}}{\rho^{1/2}}  \right).
\end{equation}
The leading order contribution in the far-field is
\begin{equation}
u^{3}_{\alpha}, \, u^{\alpha}_3  = \mathcal{O}\left( \frac{x_{\alpha} \, e^{-\rho y_1/2}}{\rho^{3/2}} \right).
\end{equation}
When $j = \beta$ and $k = \alpha$ where $\alpha \, , \beta \in [1 \, , \, 2]$, the leading  far-field contribution arises from $\gamma_{\epsilon}$, 
\begin{align}
    u^{\alpha}_{\beta} &= z(2-z) \left[ \frac{3}{ \rho^2}\left( \delta_{\alpha\beta} - \frac{2x_{\alpha}x_{\beta}}{\rho^2} \right) \right]. \label{eq:sourcedipolefarfieldform}
\end{align}

\section{Vertical Stokeslet Near the Free Surface Boundary}\label{VerticalStokesletAppendix}
Here we find the leading term in \eqref{eq:horizontalcomponentverticalStokeslet} for 
the horizontal flow field at $(\rho,0,h)$ produced by a vertical Stokeslet located at $(0,0,h)$ in the limit that $\epsilon = 1 - h \ll 1$. From \eqref{eq:fullverticalstokeslet} we have
\begin{equation}
w^3_1 = -4\int_{\gamma} \frac{k X}{(\cosh{k})^2} \frac{H^1_1(k \rho)}{\sinh{2k} - 2k} dk,
\end{equation}
where
\begin{align}
X =& h \cosh^2{k} \sinh^2{hk} + k \sinh{hk} \cosh{hk} + h^2 k \cosh^2{k}\left( \sinh{hk} \cosh{hk} + \cosh{k}\sinh{k(1-2h)} \right) \nonumber \\
&- \cosh{k}\sinh{hk}\sinh{k(1+h)} - h\cosh^2{k}\sinh{k}\sinh{k(1-2h)}+hk\cosh{k}\sinh{k(1-2h)},
\nonumber \\
\simeq & \epsilon \left(k^2 (1 + \cosh^2{k}) - 2 \cosh^2{k}\sinh^2{k}\right) + \mathcal{O}(\epsilon^2).
\end{align}
The leading order term of $u^3_1$ is that from $w^3_1$ which is the sum of the residues at the first two roots in the upper half plane to the equation $\sinh{2k} = 2k$ i.e. $k_0^{+}$ and $k_{0}^{-}$ where
\begin{equation}
2 k_{0}^{\pm} = \pm x_1 + i y_1 = 2.769 + 7.498i. 
\end{equation}
Hence, using the residue theorem we have
\begin{align}
u^3_1 &= - 4 \pi i \left( \Sigma_{k_0 \in [k_0^{+}, \, k_0^{-}]} \lim_{k \to k_0} \left( \frac{k - k_0}{\sinh{2k}-2k} \right) \frac{k H_1^1(k \rho)}{\cosh^2{k}}X \right) \nonumber \\
&= - 4 \pi i \left( \Sigma_{k_0 \in [k_0^{+}, \, k_0^{-}]} \lim_{k \to k_0} \left( \frac{1}{4 \sinh^2{k}} \right) \frac{k H_1^1(k \rho)}{\cosh^2{k}}\epsilon k^2\sinh^2{k} \right) \nonumber \\
&= -\epsilon \pi i \left( k_0^{+} \sinh^2{k_0^{+}}H_1^1(k_0^{+}\rho) + k_0^{-} \sinh^2{k_0^{-}}H_1^1(k_0^{-}\rho)  \right). \label{eq:volvoxmess1}
\end{align}
However, recall the standard result (e.g. see equation 9.2.3 of \cite{Abramowitz70}) that 
\begin{equation}
H_1^1 (z) \sim \frac{2}{\pi z} e^{i(z - 3\pi/4)} \quad \mbox{when} \quad |z| \rightarrow \infty \quad \mbox{and} \quad -\pi < \arg{z} < 2\pi.
\end{equation}
Hence, (\ref{eq:volvoxmess1}) simplifies to become
\begin{align}
u^3_1 &= -\frac{\sqrt{\pi} \epsilon}{\sqrt{\rho}}(1 - i) e^{-\rho y_1/2} \left( k_0^{+} \sinh^2{k_0^{+}}e^{i \rho x_1/2} + k_0^{-} \sinh^2{k_0^{-}}e^{-i \rho x_1/2} \right), \nonumber \\
&= -\frac{\sqrt{\pi} \epsilon}{\sqrt{\rho}} e^{- \rho y_1/2}(g + i g^{\star})(1 - i) = -\frac{2\sqrt{\pi} \epsilon}{\sqrt{\rho}} e^{- \rho y_1/2}(\mathbb{R}e(g) + \mathbb{I}m(g)), \nonumber \\ 
\end{align}
where $g$ satisfies
\begin{equation}
g = \sqrt{k_0^{+}} \sinh^2{k_0^{+}} e^{i \rho x_1/2} = e^{i \rho x_1/2} (-2.782 + 7.1238i).
\end{equation}
Rearranging this expression gives the relation in (\ref{eq:complexasymptoticresult}), namely
\begin{equation}
u^3_{\alpha} = \frac{A \epsilon x_{\alpha}}{\rho^{3/2}} e^{-\rho y_1/2}\sin{(x_1(\rho - \rho_0)}/2),
\end{equation}
where $A = 38.340$ and $\rho_0 = 0.298$. 
\end{widetext}



\end{document}